\documentclass[11pt,a4paper]{article}
\pdfoutput=1
\usepackage{jheppub}

\newcommand{\be}{\begin{equation}}
\newcommand{\ee}{\end{equation}}
\newcommand{\bea}{\begin{eqnarray}}
\newcommand{\eea}{\end{eqnarray}}

\newcommand{\eqn}[1]{(\ref{#1})}

\newcommand{\mt}[1]{\textrm{\tiny #1}}
\def\nc {N_\mt{c}}

\newcommand{\sac}{\, , \qquad}

\newcommand{\uh}{u_\mt{H}}

\newcommand{\cf}{{\cal F}}
\newcommand{\cb}{{\cal B}}
\newcommand{\ch}{{\cal H}}
\newcommand{\cn}{{\cal N}}
\newcommand{\cl}{{\cal L}}

\newcommand{\vp}{\varphi}

\title{Drag force in a strongly coupled anisotropic plasma}
\author[a,b]{Mariano Chernicoff,}
\author[a]{Daniel Fern\'andez,}
\author[a,c]{David Mateos,}
\author[d,e]{and Diego Trancanelli}
\affiliation[a]{Departament de F\'\i sica Fonamental \&  Institut de 
Ci\`encies del Cosmos (ICC), Universitat de Barcelona (UB), Mart\'{\i}  i
Franqu\`es 1, E-08028 Barcelona, Spain}  
\affiliation[b]{Department of Applied Mathematics and Theoretical Physics, Centre for Mathematical Sciences,
Wilberforce Road, Cambridge, CB3 0WA, UK}
\affiliation[c]{Instituci\'o Catalana de Recerca i Estudis Avan\c cats (ICREA),
Passeig Llu\'\i s Companys 23, E-08010, Barcelona, Spain} 
\affiliation[d]{Department of Physics, University of Wisconsin, Madison, WI
53706, USA}  
\affiliation[e]{Instituto de F\'\i sica, Universidade de S{\~a}o Paulo, 05314-970 S{\~a}o Paulo, Brazil}

\date{\today}

\abstract{We calculate the drag force experienced by an infinitely massive quark propagating at constant velocity through an anisotropic, strongly coupled ${\cal N}=4$ plasma by means of its gravity dual. We find that the gluon cloud trailing behind the quark is generally misaligned with the quark velocity, and that the latter is also misaligned with the force. The drag coefficient $\mu$ can be larger or smaller than the corresponding  isotropic value depending on the velocity and the direction of motion. In the ultra-relativistic limit we find that  generically $\mu \propto p$. We discuss the conditions under which this behaviour may extend to more general situations.} 
  

\emailAdd{M.Chernicoff@damtp.cam.ac.uk}
\emailAdd{daniel@ffn.ub.edu} 
\emailAdd{dmateos@icrea.cat} 
\emailAdd{dtrancan@fma.if.usp.br} 

\begin{document}

\begin{flushright}
DAMTP-2012-12 \\
ICCUB-12-088 \\
MAD-TH-12-01
\end{flushright}

\maketitle
\setlength{\parskip}{8pt}
\section{Introduction}
\label{intro}
A remarkable conclusion from the experiments at the Relativistic Heavy Ion Collider (RHIC)  \cite{rhic,rhic2} and at the Large Hadron Collider (LHC) \cite{lhc} is that the quark-gluon plasma (QGP) does not behave as a weakly coupled gas of quarks and gluons, but rather as a strongly coupled fluid \cite{fluid,fluid2}. This renders perturbative methods inapplicable in general. The lattice formulation of Quantum Chromodynamics (QCD) is also of limited utility, since for example it is not well suited for studying real-time phenomena. This has provided a strong motivation for understanding the dynamics of strongly coupled non-Abelian plasmas through the gauge/string duality \cite{duality,duality2,duality3} (see \cite{review} for a recent review of applications to the QGP). 

For a period of time $\tau_\mt{out}$ immediately after the collision, the system thus created is far from equilibrium. After a time $\tau_\mt{iso} > \tau_\mt{out}$ the system becomes locally isotropic and a standard hydrodynamic description becomes applicable. It has been proposed than an intrinsically anisotropic hydrodynamical description can be used to describe the system at intermediate times $\tau_\mt{out} < \tau < \tau_\mt{iso}$
\cite{ani,ani2,ani3,ani4,ani5,ani6,ani7,ani8,ani9}. In this phase the plasma is assumed to have significantly unequal pressures in the longitudinal and transverse directions. The standard hydrodynamic description is a derivative expansion around equal pressures, and therefore it is not applicable in this regime. In contrast, the intrinsically anisotropic hydrodynamical description is a derivative expansion around an anisotropic state, and hence in this case the requirement that derivative corrections be small does not imply small pressure differences. In a real collision the degree of anisotropy will decrease with time, but for some purposes it is a good approximation to take it to be constant over an appropriate time scale. 

Motivated by these considerations, in this paper we will investigate the effect of an intrinsic anisotropy on  the drag force felt by an infinitely massive quark propagating through a strongly coupled plasma. For this purpose we will examine a string moving in a gravity solution \cite{prl,jhep} dual to an anisotropic ${\cal N}=4$ super Yang-Mills plasma. As we will review below, the plasma is held in anisotropic equilibrium by an external force. The gravity solution possesses an anisotropic horizon, it is completely regular on and outside the horizon, and it is solidly embedded in type IIB string theory. For these reasons it provides an ideal toy model in which questions about  anisotropic effects at strong coupling can be addressed from first principles.

We will pay particular attention to the ultra-relativistic behavior of the drag force, which can be determined analytically. To avoid confusion, we emphasize from the beginning that our results correspond to sending the quark mass to infinity first, and then sending $v\to 1$. In particular, this means that in any future attempt to connect our results to the phenomenology of the QGP, this connection can only be made to the phenomenology of heavy quarks moving through the plasma.

Following the original calculations \cite{drag1,drag2} of the drag coefficient, the closely related diffusion coefficient was obtained independently in \cite{CasalderreySolana:2006rq}. These seminal papers have been generalized and elaborated on in a vast number of subsequent contributions \cite{Herzog:2006se}, including in particular comparisons with the corresponding weakly-coupled results \cite{Chesler:2006gr}, as well as extensive analyses of the energy-momentum tensor which provide  a
detailed picture of the directionality of energy flow away from the moving quark \cite{Friess:2006aw}. Examples of holographic studies of the drag force in the presence of anisotropies and/or inhomogeneities include \cite{schalm,panigrahi}.

\section{Gravity solution}
\label{grav}
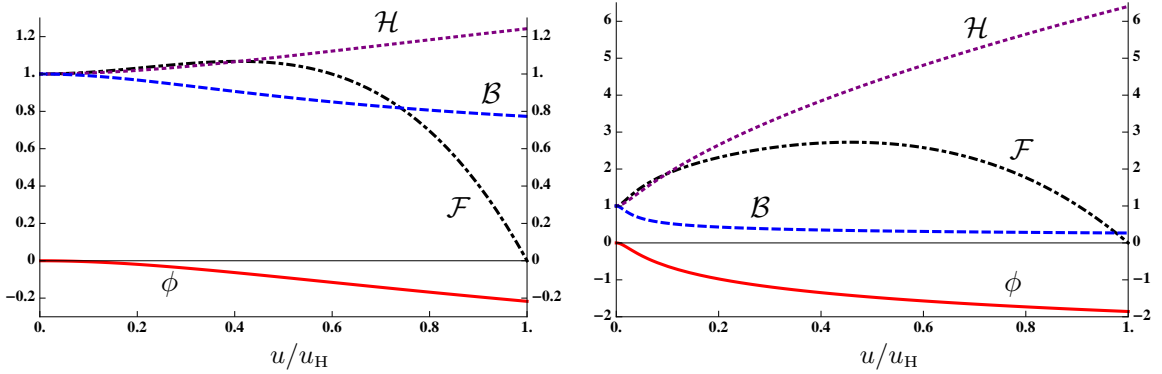
\begin{figure}[tb]
\begin{center}
\begin{tabular}{cc}
\includegraphics[scale=0.8]{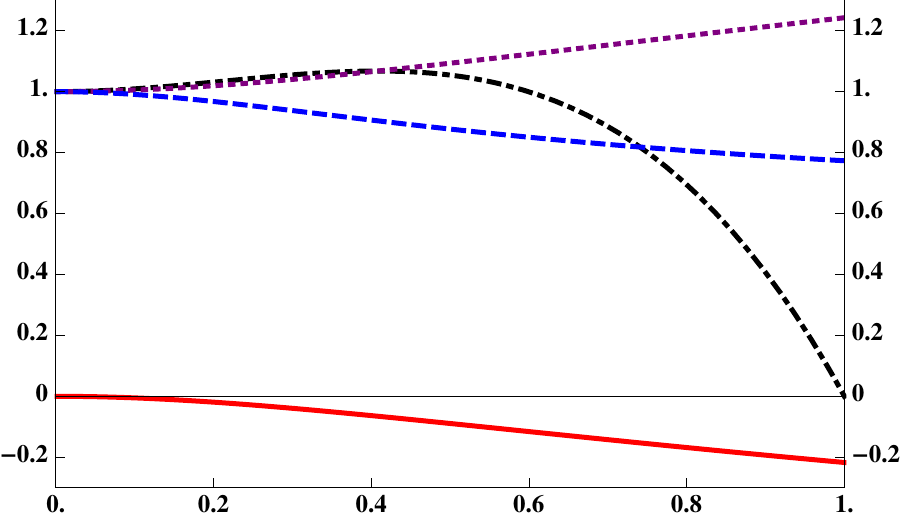}
\put(-109,-10){\small $u/\uh$}
\put(-109,-14){$$}
\put(-150,17){$\phi$}
\put(-70,115){$\ch$}
\put(-30,88){$\cb$}
\put(-43,45){$\cf$}
&
\includegraphics[scale=0.8]{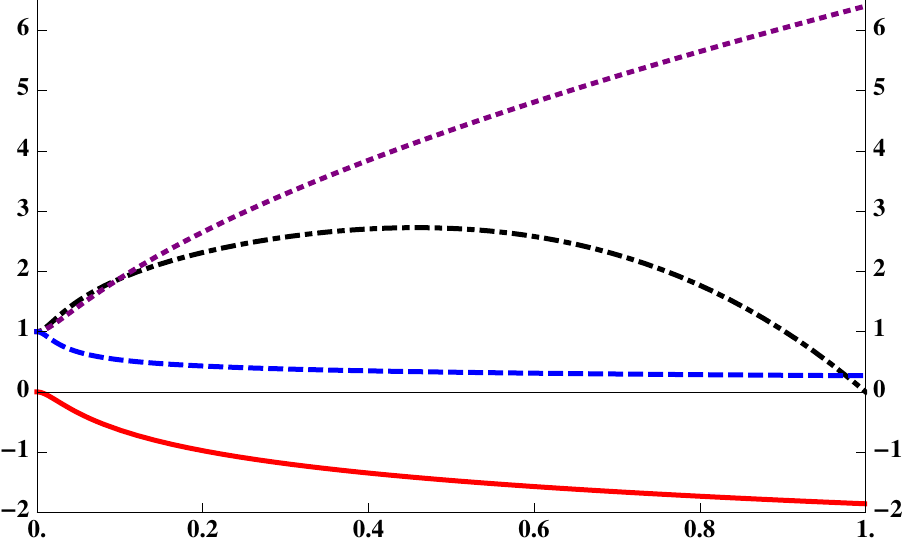}
\put(-109,-10){\small $u/\uh$}
\put(-109,-14){$$}
\put(-55,17){$\phi$}
\put(-70,113){$\ch$}
\put(-150,45){$\cb$}
\put(-53,67){$\cf$}
\end{tabular}
\caption{\small Metric functions for $a/T\simeq 4.4$ (left) and $a/T\simeq 86$ (right).
\label{plots}
}
 \end{center} 
 \end{figure}

The type IIB supergravity solution of \cite{prl,jhep} in the string frame takes the form
\bea
&&\hskip -.35cm 
ds^2 =  \frac{L^2}{u^2}
\left( -\cf \cb\, dt^2+dx^2+dy^2+ \ch dz^2 +\frac{ du^2}{\cf}\right) +
L^2 e^{\frac{1}{2}\phi} d\Omega_5^2, 
\,\,\,\,\,\,\label{sol1} \\
&& \hskip -.35cm \chi = az \sac \phi=\phi(u) \,,
\label{sol2}
\eea
where $\chi$ and $\phi$ are the axion and the dilaton, respectively, and $(t,x,y,z)$ are the gauge theory coordinates. 
Since there is rotational invariance in the $xy$-directions, we will refer to these as the transverse directions, and to $z$ as the longitudinal direction. $\cf, \cb$ and $\ch$ are functions of the holographic radial coordinate $u$ that were determined numerically in \cite{prl,jhep}. Their form for two values of $a/T$ is plotted in Fig.~\ref{plots}. 
The horizon lies at $u=\uh$, where $\cf=0$, and the boundary at $u=0$, where $\cf=\cb=\ch=1$ and $\phi=0$. The metric near the boundary asymptotes to $AdS_5 \times S^5$. Note that the axion is linear in the $z$-coordinate. The proportionality constant $a$ has dimensions of mass and is a measure of the anisotropy. The axion profile is dual in the gauge theory to a position-dependent theta parameter of the form $\theta \propto z$. This acts as an isotropy-breaking external source that forces the system into an anisotropic equilibrium state. 

\begin{figure}[t!]
\begin{center}
\includegraphics[scale=0.85]{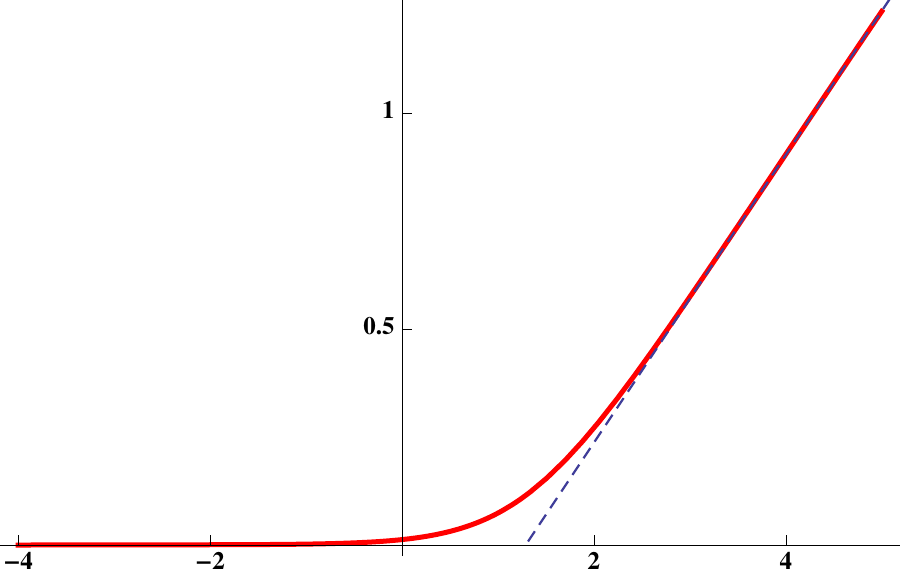}
 \begin{picture}(0,0)
   \put(1,8){\small{$\log \left( 
   {\displaystyle \frac{a}{T}} \right)$}}
    \put(-120,130){\small{$\log \left( 
    {\displaystyle \frac{s}{s_\mt{iso}}} \right)$}}
 \end{picture}
\caption{\small Log-log plot of the entropy density as a function of 
$a/T$, with $s_\mt{iso}$ defined as in eqn.~\eqn{siso}. The dashed blue line is a straight line with slope $1/3$.
\label{scalings}}
\end{center}
\end{figure}
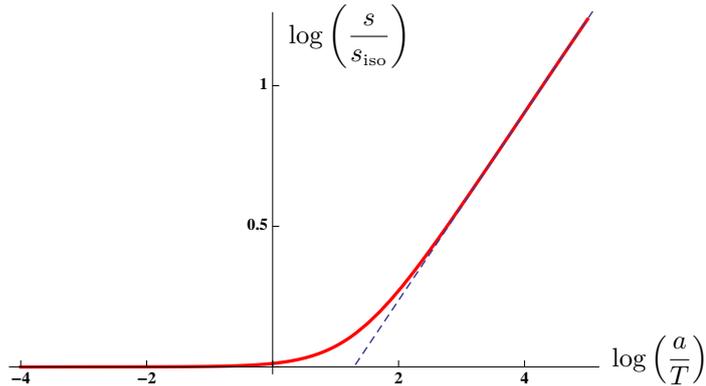

If $a=0$ then the solution reduces to the isotropic black D3-brane solution dual to the isotropic 
$\cn=4$ theory at finite temperature. In this case
\be
\cb=\ch=1 \sac \chi=\phi=0 \sac \cf = 1-\frac{u^4}{\uh^4}
\sac \uh = \frac{1}{\pi T} 
\label{iso}
\ee
and the entropy density takes the form 
\be
s_\mt{iso}= \frac{\pi^2}{2} \nc^2 T^3 \,.
\label{siso}
\ee
Fig.~\ref{scalings} shows the entropy density of the anisotropic
plasma as a function of the dimensionless ratio $a/T$, normalized to the entropy density of the isotropic plasma at the same temperature. At small $a/T$ the entropy density scales as in the isotropic case, whereas at large $a/T$ it scales as \cite{ALT,prl,jhep}
\be
s = c_\mt{ent} \nc^2 a^{1/3} T^{8/3} \,,
\label{larges}
\ee
where $c_\mt{ent}$ is a constant that can be determined numerically. 

A feature of the solution \eqn{sol2} that played an important role in the analysis of \cite{prl,jhep} is the presence of a conformal anomaly. Its origin lies in the fact that diffeomorphism invariance    in the radial direction $u$ gets broken in the process of renormalization of the on-shell supergravity action. In the gauge theory this means that scale invariance is broken by the renormalization process. One manifestation of the anomaly is the fact that, unlike the entropy density, other thermodynamic quantities do not depend solely on the ratio $a/T$ but on $a$ and $T$ separately. Fortunately, this will not be the case for our drag force, which will take the form $F(a,T) = T^2 f$ with $f$ a function of the ratio $a/T$ alone. The reason for this is that no regularization procedure is necessary for the computation of the drag force, and thus diffeomorphism invariance is preserved. We will also verify this analytically in certain limits, and numerically for general values of $a$ and $T$.

\section{Drag force}
\label{dragsec}
Extending the isotropic analysis of Refs.~\cite{drag1,drag2}, in this section we will consider the drag force acting on an infinitely massive quark moving at constant velocity through the anisotropic $\cn=4$ plasma described by \eqn{sol2}. A simple model for this system is described by the equation of motion 
\be
\frac{d \vec p}{dt} = -\mu \vec p +\vec F \,,
\ee
where $\vec p$ is the quark's momentum, $\mu$ is a drag coefficient, and $\vec F$ is an external force. The necessary force to keep a steady motion is $\vec F = \mu \vec p$. An observation that will be important for us is that, in the case of an anisotropic medium, the drag coefficient is not just a number but  a matrix. In our case we will see that this matrix is diagonal, $\mu = \mbox{diag} (\mu_x, \mu_y, \mu_z)$ with $\mu_x=\mu_y\neq \mu_z$. Thus we should expect that the force and the momentum or the velocity of the quark will not be aligned in general, and indeed our calculations will reproduce  this feature. We will also see that, unlike in \cite{drag1,drag2}, the drag coefficient is momentum-dependent. 

On the gravity side the quark is described by a string propagating 
in the background \eqn{sol2}. The string action is 
\be
S =- \frac{1}{2\pi\alpha'}\int d^2\sigma\,  \sqrt{-g} = 
\int d^2\sigma \, {\cal L}  \,,
\label{stringaction}
\ee
where $g$ is the induced worldsheet metric. With the $L^2$ factor from the
spacetime metric the Lagrangian scales as $L^2/2\pi
\alpha'=\sqrt{\lambda}/2\pi$. We will set this factor to one in
intermediate  expressions, and we will reinstate it at the end.  Denoting the spacetime coordinates collectively by $X^M$, the  flow of spacetime momentum $\Pi_M$ along the string is given by
\be
\Pi_M = \frac{\partial \cl}{\partial (\partial_\sigma X^M)} \,.
\ee
Physically, one can imagine that the  external force on the quark needed to sustain steady motion may be exerted by attaching the endpoint of the string to a D7-brane and turning on a constant electric field $F_{MN}=\partial_{[M}A_{N]}$ 
on the brane. In other words, we add to the action \eqn{stringaction} the boundary term 
\be
S_\mt{bdry} = -\int_{\partial \Sigma}  d\tau A_N \partial_\tau X^N =
-\frac{1}{2} \int_{\partial \Sigma} d\tau F_{MN} X^M \partial_\tau X^N \,.
\ee
Demanding that the boundary term arising from variation of the total action $S+S_\mt{bdry}$ vanish yields the boundary condition
\be
\left. \Pi_M + F_{MN} \partial_\tau X^N \right| _{\partial \Sigma} =0\,.
\label{bc}
\ee

We will now specify to the case of a quark moving steadily through the plasma. The string will not move along the sphere directions, so this part of the metric will  play no role in the following. Also, given the rotational symmetry in the $xy$-directions, we will assume that $y=0$. We fix the static gauge by identifying $(t,u)=(\sigma^0,\sigma^1)$ and consider a string embedding of the form 
\bea
x(t,u)&\to&  \Big( vt + x(u) \Big) \sin \varphi \,, \\
z(t,u)&\to&  \Big( vt + z(u) \Big) \cos \varphi \,, 
\label{pro}
\eea
corresponding to a quark moving with velocity $v$ in the $xz$-plane at an angle $\varphi$ with the $z$-axis. Under these circumstances the Lagrangian takes the form 
\be
\cl = - \left[ \frac{\cb \cf +  \sin^2 \varphi \, ( \cb \cf^2 x'^2 -v^2)
+ 
\ch \cos^2 \varphi \, \Big[ \cb \cf^2 z'^2 -v^2 - \cf v^2 (x' - z')^2 \sin^2
\varphi \Big]}{\cf u^4}  \right]^{1/2} \,. 
\ee
The rates at which energy and momentum flow down the string towards the horizon are then
\bea
- \Pi_t &=& \frac{1}{\cl u^{4}} \,\cb \cf v \Big[  x' \sin^2 \vp + \ch z' \cos^2 \vp \Big] 
\,,\nonumber \\ [1.3mm]
\Pi_x &=& \frac{1}{\cl u^{4}} \,\Big[ \cb \cf \,x' + \ch v^2 (z' -x') \cos^2 \varphi \Big] \sin \varphi  \,,\nonumber \\ [1.3mm]
\Pi_z &=& \frac{1}{\cl u^{4}} \, \ch \Big[  \cb \cf\,  z' + v^2 (x' - z') \sin^2 \varphi \Big] \cos \varphi   \,,
\label{momenta}
\eea
where $'$ denotes differentiation with respect to $u$, and the boundary conditions \eqn{bc}  become
\be
\Pi_x = F_x \sac \Pi_z=F_z \sac - \Pi_t = F_x \, v \sin \vp + F_z \, v \cos \vp \,,
\ee
where $(F_x,F_z)$ denote the components of the external force (the electric field). The first two equations are the statement that the external force exactly compensates for the momentum lost by the quark into the medium. The third equation is identically satisfied by virtue of \eqn{momenta}, and it expresses the fact that the work done by the external force precisely equals the rate at which  the quark deposits energy into the medium.  As we will see below, the energy and the  momentum flow from the boundary to the horizon (i.e.~$\Pi_x, \Pi_z$ and $-\Pi_t$) are positive provided the string trails behind the quark (i.e.~if  $x', z'$ are negative), as we would expect on physical grounds. This can be easily seen by inspection in the simple cases of motion along the 
$z$-direction ($\vp=0$), for which  
\be
\Pi_z = -\frac{\cb \cf \ch \, z'}{u^2 \sqrt{\cb - \frac{\ch v^2}{\cf} + \cb \cf \ch \, z'^2}}
 \sac -\Pi_t = \Pi_z \, v \sac \Pi_x=0 \,,
\ee
and of motion along the $x$-direction ($\vp=\pi/2$), for which  
\be
\Pi_x =  -\frac{\cb \cf  \, x'}{u^2 \sqrt{\cb - \frac{ v^2}{\cf} + \cb \cf  \, x'^2}}
 \sac -\Pi_t = \Pi_x \, v \sac \Pi_z = 0 \,.
\ee

We will now determine the string profile and the corresponding values of the energy and momentum flows for arbitrary $v, \vp$. The first observation is that, generically, the string does not trail  `below' its endpoint's trajectory. In other words, $x(u)\neq z(u)$. Indeed, if  $x(u)=z(u)$ then the ratio of the momenta would be given by
\be
\frac{\Pi_x}{\Pi_z}= \frac{\tan \varphi}{H(u)} \,,
\ee
which would be a contradiction because the left-hand side is constant whereas the right-hand side is not. In order to determine the correct string profile we invert the relations \eqn{momenta} to find
\be
x' =\pm \frac{\ch v}{\cf \sqrt{\cb \ch}} \frac{N_x}{\sqrt{N_x N_z-D}} \sac
z' = \pm \frac{v}{\cf \sqrt{\cb \ch}} \frac{N_z}{\sqrt{N_x N_z-D}}\,,
\label{primes}
\ee
where
\bea
N_x&=& -\Pi_x (\cb \cf  \csc \vp -  v^2 \sin \vp)+\Pi_z v^2 \cos \vp \,, \\[1.7mm]
N_z&=& -\Pi_z (\cb \cf  \sec \vp - \ch v^2 \cos \vp) + \Pi_x \, \ch v^2 \sin \vp 
\,, \\[1.7mm]
D&=& \frac{\cb \cf  \csc \vp \sec \vp}{u^4} 
\Big[  \Pi_x \Pi_z u^4 - \ch v^2 \cos \vp \sin \vp \Big] 
\Big[ \cb \cf - v^2 \left( \ch \cos^2 \vp + \sin^2 \vp \right) \Big] \,.\,\,\,\,\,\,\,\,\,\, 
\label{nnd}
\eea
The factor $N_x N_z -D$ inside the square root in the denominator of \eqn{primes} is positive at the boundary, where $\cb, \cf, \ch \to 1$ and $u\to 0$, and also at the horizon, where $\cf \to 0$, and generically it becomes negative in some region in between. In other words, it vanishes at two different values of $u$ between the boundary and the horizon. To see this, consider the last factor in square brackets in \eqn{nnd}. $\cb \cf$ ($\ch$) is monotonically decreasing (increasing) from the boundary to the horizon, so this factor is positive at the boundary and negative at the horizon. Therefore there exists a critical value $u_c$ in between such that 
\be 
 \cb_c \, \cf_c - v^2 \left( \ch_c \cos^2 \vp + \sin^2 \vp \right) = 0\,,
 \label{hc}
 \ee
where $\cb_c=\cb(u_c)$, etc. At this point $D=0$ and 
\be
\left. N_x N_z \right|_{u_c} = 
- v^4 \left( \ch_c\,  \Pi_x \cos \vp - \Pi_z \sin \vp \right)^2 \,,
\ee
which is negative unless the momenta are related through 
\be
\frac{\Pi_x}{\Pi_z} = \frac{\tan \vp}{\ch_c}\,.
\label{relation}
\ee
If this condition is not satisfied then $N_x N_z - D$ is negative in some interval $u_1 <u_c < u_2$ and vanishes at $u=u_1$ and at $u=u_2$. This type of solutions correspond to strings with two endpoints at the boundary. Here we wish to study isolated quarks, which are described by strings that extend all the way from the boundary to the horizon, so we must require that $N_x N_z - D$ is non-negative for all  $0< u< \uh$. This is satisfied if and only if \eqn{relation} holds and if the two zeros of $D$ coincide with one another, i.e.~if the first square bracket in \eqn{nnd} also vanishes at $u=u_c$. The latter condition, together with \eqn{relation}, allows us to solve for the two momenta independently with the final result:
\be
\Pi_x =  \frac{v \sin \vp}{u_c^2} \sac 
\Pi_z =  \ch_c\,  \frac{v  \cos \vp}{u_c^2} \,.
\label{result}
\ee
Under these circumstances the denominator in \eqn{primes} is always real and positive except at $u_c$, where it vanishes. At this point the numerators also vanish and the functions $x', z'$ are  smooth and negative for all $0< u< \uh$ provided in \eqn{primes} we choose the positive sign for $u<u_c$ and the negative sign for $u>u_c$.

In summary, we have obtained the force $\vec F = (\Pi_x, \Pi_z) $ that must be exerted on the quark in order to maintain its stationary motion,
\be
\vec F = \frac{\sqrt{\lambda}}{2\pi}\, 
\frac{v}{u_c^2} \, (\sin \vp, \ch_c \cos \vp) \,,
\label{dragv}
\ee
in terms of the quark's velocity $\vec v=v(\sin \vp, \cos\vp)$. (In this equation we have reinstated the factor $L^2/2\pi\alpha'$.) 
The external force $\vec F$ is equal to minus the drag force exerted on the quark by the plasma, but in a slight abuse of language we will refer to $\vec F$ itself as the drag force.
Note that $\vec v$ and $\vec F$ are not aligned with one another except in the isotropic case, for which $\ch_c = 1$, or in the cases in which the velocity is aligned with one of the axes, in which $\varphi=0, \pi/2$. Note also the force depends on the velocity both through the explicit factors of $v$ and $\varphi$ in eqn.~\eqn{dragv} and implicitly through the value of  $\ch_c$, which is a solution of the $\vec v$-dependent equation \eqn{hc}.

Substituting the result \eqn{result} in \eqn{primes} we obtain the form of the
string profile as a function of the velocity. The projection of this profile on the $xz$-plane has tangent vector $\vec \tau = (\tau_x,\tau_z)=(x' \sin \vp, z'
\cos\vp)$. The angle $\vp_\tau$ between this vector and the $z$-axis is 
\be
\tan \vp_\tau = \frac{\tau_x}{\tau_z} =\epsilon \tan \vp \sac \epsilon = 1 + \frac{\cb \cf (\ch -\ch_c)}{\cb \cf \ch_c -\cb_c \cf_c \ch} \,.
\label{phitau}
\ee
At the horizon we have $\cf=0$ and thus $\epsilon=1$, which means that deep in the infrared the
string aligns itself with the velocity. However, near the boundary $\cb, \ch,
\cf \to 1$ and thus 
\be
\epsilon \to 1 + \frac{1 -\ch_c}{\ch_c - \cb_c \cf_c } \,.
\ee
This is different from unity for generic $\vp$, and so the string does not align itself with the velocity except if $\vp = 0$ or $\vp=\pi/2$. In these two special cases the entire string profile (not just the infrared part) aligns itself with the $z$- or the $x$-axis, respectively, because $\epsilon$ remains finite in these limits whereas $\tan \vp \to 0, \infty$, respectively. Note also that the vector $\vec \tau$ is not aligned with the force $\vec F$ either, since $\epsilon\neq \ch_c^{-1}$.

The formulas above reduce to the correct expressions in the isotropic limit \eqn{iso}. In this case eqn.~\eqn{hc} yields 
\be
u_c^2 = \uh^2 \sqrt{1-v^2} = \frac{\sqrt{1-v^2}}{\pi^2 T^2} 
\label{uiso}
\ee
and the force \eqn{dragv} becomes 
\be
\vec F_\mt{iso}(T) = F_\mt{iso}(T) (\sin \vp, \cos \vp) 
\ee
with 
\be
F_\mt{iso}(T) = \frac{\pi}{2}\, \sqrt{\lambda}\, T^2 \frac{v}{\sqrt{1-v^2}}\,,
\label{fisoT}
\ee
as in \cite{drag1,drag2}.  For later purposes it is useful to rewrite this result as 
\be
F_\mt{iso}(s) = 
 \frac{\sqrt{\lambda} \, s^{2/3}}{ (2\pi)^{1/3}\nc^{4/3}} \, 
\frac{v}{\sqrt{1-v^2}} 
\label{fisos}
\ee
in terms of the entropy density \eqn{siso} of the isotropic $\cn=4$ plasma.

\section{Results} 
\label{results}
\begin{figure}
\begin{center}
\begin{tabular}{cc}
\setlength{\unitlength}{1cm}
\hspace{-0.9cm}
\includegraphics[width=7cm]{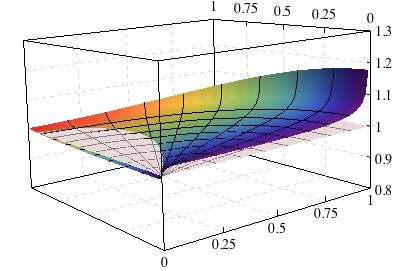} 
\qquad\qquad & 
\includegraphics[width=7cm]{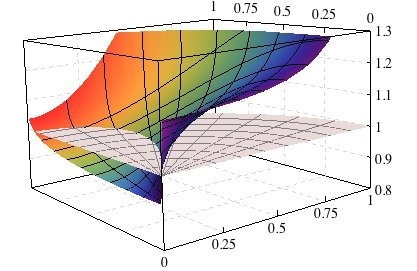}
\qquad
\put(-243,55){\rotatebox{90}{$F/F_\mt{iso}(T)$}} 
\put(-6,55){\rotatebox{90}{$F/F_\mt{iso}(T)$}} 
\put(-53,10){$v_z$}
\put(-290,10){$v_z$}
\put(-295,135){$v_x$}
\put(-59,133){$v_x$}
\\
(a) & (b)\\
& \\
\hspace{-0.9cm}
\includegraphics[width=7cm]{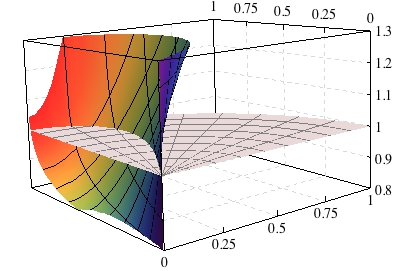} 
\qquad\qquad & 
\includegraphics[width=7cm]{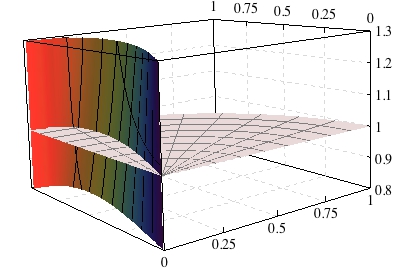}
\qquad
\put(-243,55){\rotatebox{90}{$F/F_\mt{iso}(T)$}} 
\put(-6,55){\rotatebox{90}{$F/F_\mt{iso}(T)$}} 
\put(-53,10){$v_z$}
\put(-290,10){$v_z$}
\put(-295,135){$v_x$}
\put(-59,133){$v_x$}
\\
(c)& (d) 
\end{tabular}
\end{center}
\caption{Drag force as a function of the quark velocity 
$(v_x, v_z)=v (\sin \vp, \cos \vp )$ for a quark moving  through an anisotropic plasma with $a/T=1.38 \mbox{(a)}, 4.41 \mbox{(b)}, 12.2 \mbox{(c)}, 86 \mbox{(d)}$. $F$ is plotted in the appropriate units to facilitate comparison with the isotropic result \eqn{fisoT} for a plasma at the same temperature.}
\label{3Dtemp}
\end{figure}
\begin{figure}
\begin{center}
\begin{tabular}{cc}
\setlength{\unitlength}{1cm}
\hspace{-0.9cm}
\includegraphics[width=7cm]{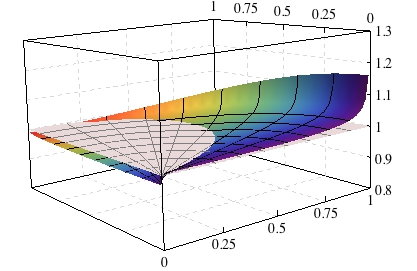} 
\qquad\qquad & 
\includegraphics[width=7cm]{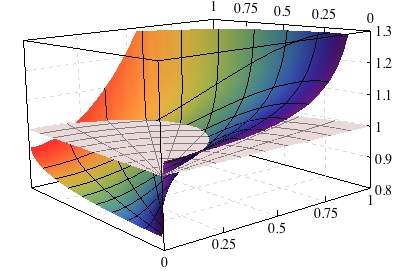}
\qquad
\put(-243,55){\rotatebox{90}{$F/F_\mt{iso}(s)$}} 
\put(-6,55){\rotatebox{90}{$F/F_\mt{iso}(s)$}} 
\put(-53,10){$v_z$}
\put(-290,10){$v_z$}
\put(-295,135){$v_x$}
\put(-59,133){$v_x$}
\\
(a) & (b)\\
& \\
\hspace{-0.9cm}
\includegraphics[width=7cm]{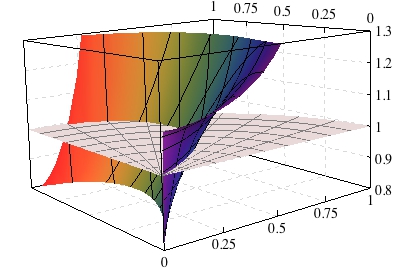} 
\qquad\qquad & 
\includegraphics[width=7cm]{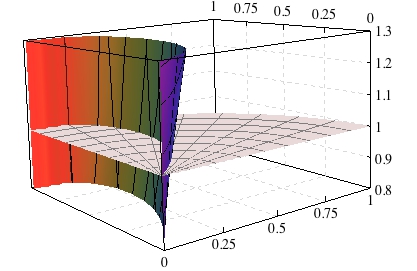}
\qquad
\put(-243,55){\rotatebox{90}{$F/F_\mt{iso}(s)$}} 
\put(-6,55){\rotatebox{90}{$F/F_\mt{iso}(s)$}} 
\put(-53,10){$v_z$}
\put(-290,10){$v_z$}
\put(-295,135){$v_x$}
\put(-59,133){$v_x$}
\\
(c)& (d) 
\end{tabular}
\end{center}
\caption{Drag force as a function of the quark velocity 
$(v_x, v_z)=v (\sin \vp, \cos \vp )$ for a quark moving  through an anisotropic plasma with $a \nc^{2/3}/s^{1/3}=0.80 \mbox{(a)}, 2.47 \mbox{(b)}, 6.24 \mbox{(c)}, 35.5 \mbox{(d)}$. $F$ is plotted in the appropriate units to facilitate comparison with the isotropic result \eqn{fisos} for a plasma at the same entropy density.}
\label{3Dentro}
\end{figure}
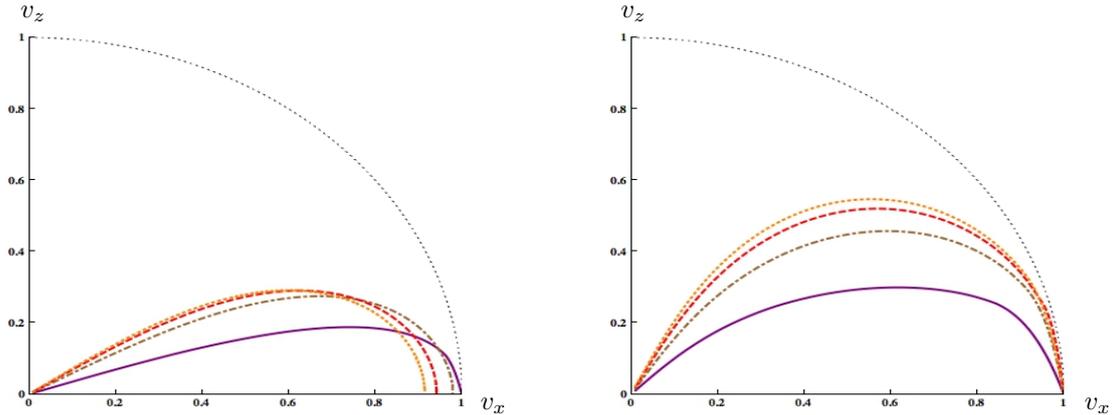
\begin{figure}[h]
\begin{center}
\setlength{\unitlength}{1cm}
\,\,\,\,\,\,
\includegraphics[width=6cm,height=5cm]
{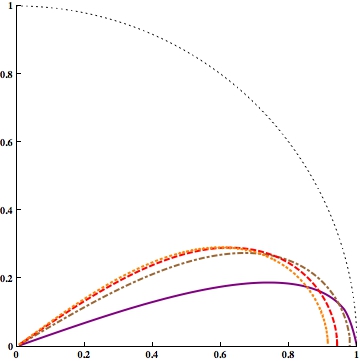}\qquad\qquad\,\,\,\,
\includegraphics[width=6cm,height=5cm]
{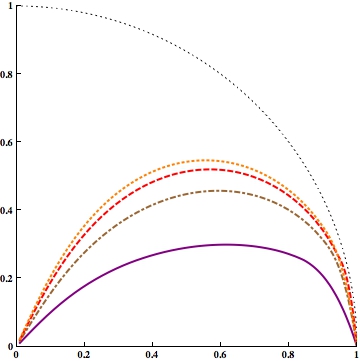}
 \begin{picture}(0,0)
   \put(0.1,0){\small{$v_x$}}
   \put(-6,5.2){\small{$v_z$}}
    \put(-7.85,0){\small{$v_x$}}
   \put(-13.9,5.2){\small{$v_z$}}
 \end{picture}
\caption{(Left) Values of the  velocity at which the drag in an anisotropic plasma with (from top to bottom) $a/T = 1.38, 4.41, 12.2, 86$ equals the drag in an isotropic plasma at the same temperature. (Right) Values of the  velocity at which the drag in an anisotropic plasma with (from top to bottom) $a\nc^{2/3}/s^{1/3} = 0.80, 2.47, 6.24, 35.5$ equals the drag in an isotropic plasma at the same entropy density. For a given value of $a/T$ or $a\nc^{2/3}/s^{1/3}$, the anisotropic drag is larger (smaller) than the isotropic drag above (below) the corresponding curve.
}
\label{inter}
\end{center}
\end{figure}

\begin{figure}[h!!]
\begin{center}
\begin{tabular}{cc}
\setlength{\unitlength}{1cm}
\includegraphics[width=6cm,height=4cm]{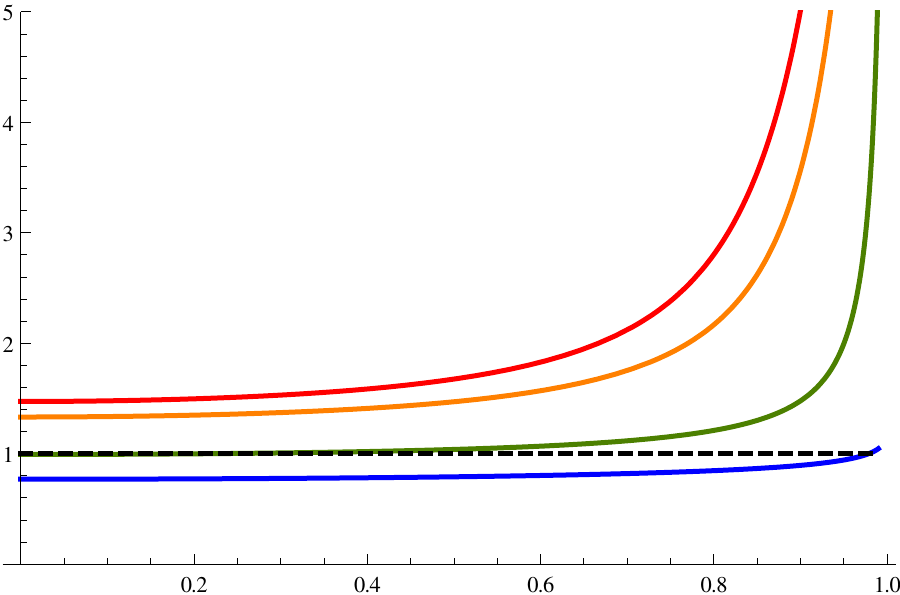} \qquad & \qquad\qquad
\includegraphics[width=6cm,height=4cm]{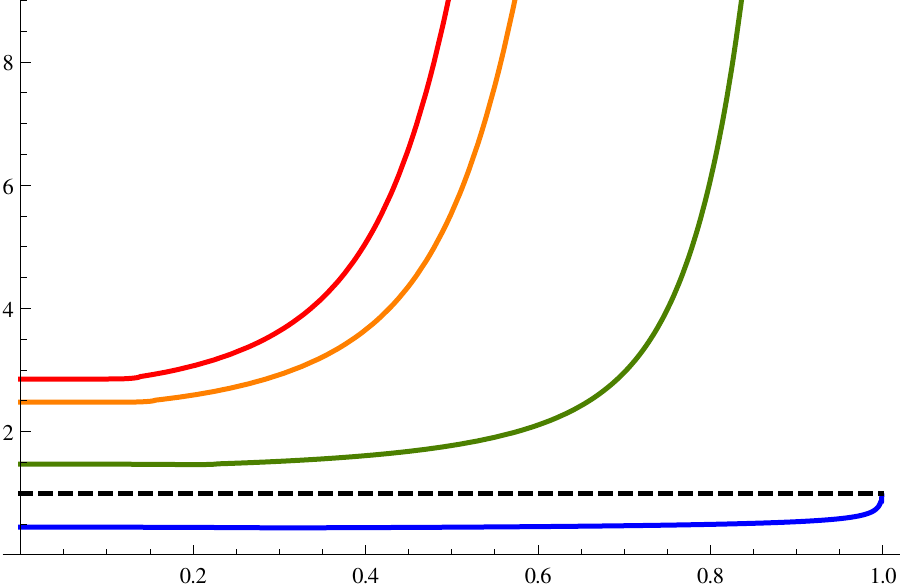}
\put(-420,45){\rotatebox{90}{$F/F_\mt{iso}(T)$}} 
\put(-190,45){\rotatebox{90}{$F/F_\mt{iso}(T)$}} 
 \\
$v$ & \qquad\qquad $v$ \\
& \\
\includegraphics[width=6cm,height=4cm]{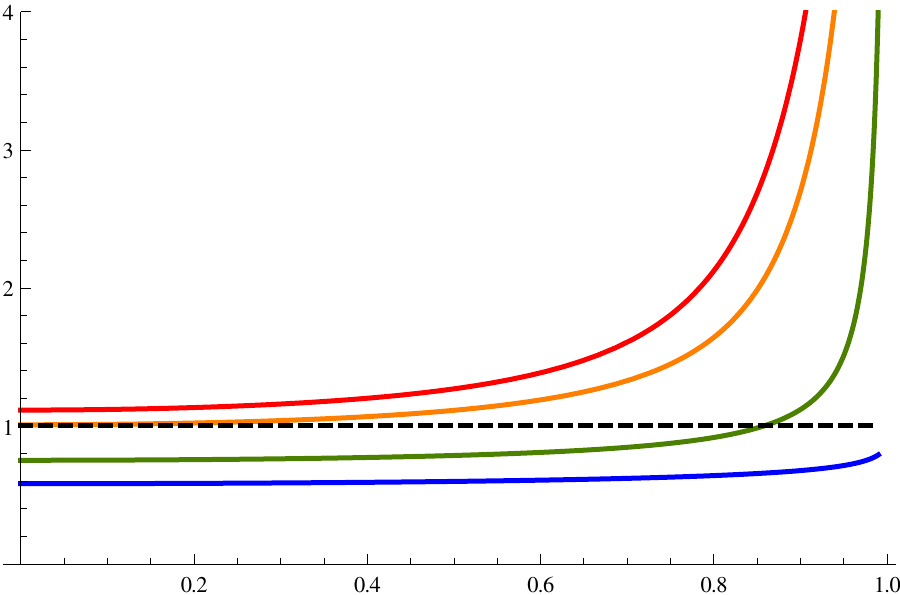} \qquad & \qquad\qquad
\includegraphics[width=6cm,height=4cm]{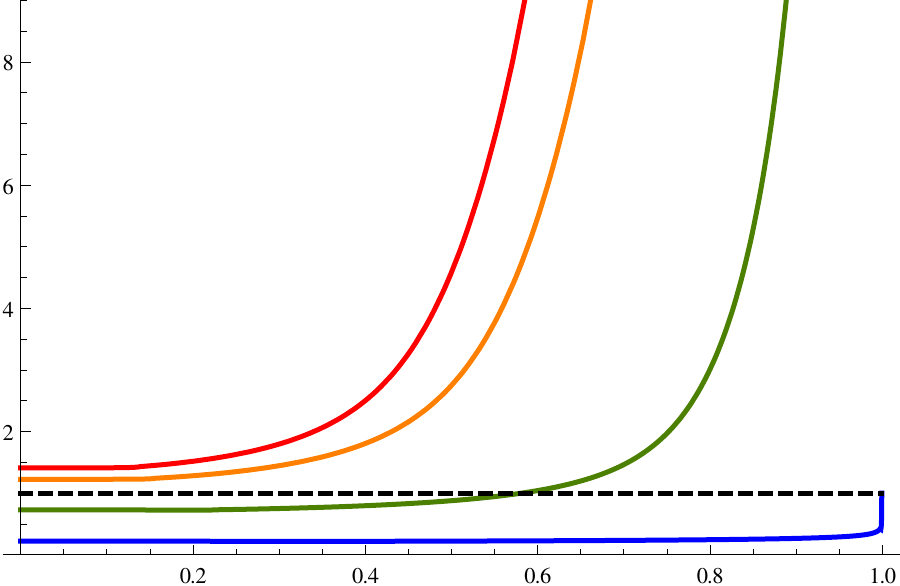}
\put(-420,43){\rotatebox{90}{$F/F_\mt{iso}(s)$}} 
\put(-190,43){\rotatebox{90}{$F/F_\mt{iso}(s)$}} 
 \\
$v$ & \qquad\qquad $v$ 
\end{tabular}
\end{center}
\caption{Drag force as a function of the velocity for a quark moving  through an anisotropic plasma with $a/T=12.2$, or equivalently $a \nc^{2/3}/s^{1/3}=6.24$,  (left column) and $a/T=86$, or equivalently $a \nc^{2/3}/s^{1/3}=35.5$, (right column) along four different directions lying at angles (curves from top to bottom) $\vp= 0, \pi/6, \pi/3, \pi/2$ with respect to the longitudinal direction $z$. $F$ is plotted in the appropriate units to facilitate comparison with the isotropic result for a plasma at the same temperature (top row) or at the same entropy density (bottom row). The isotropic result is given in eqs.~\eqn{fisoT} and \eqn{fisos}.
}
\label{dragvelo}
\end{figure}
With the groundwork above in place, we can now proceed to state our results. Since for general $a$ the metric functions in \eqn{sol2} are only known numerically, we have numerically determined the drag force as a function of the magnitude of the quark velocity $v$, of its direction $\vp$, and of the anisotropy $a$ measured in units of the temperature $T$ or in units of the entropy density $s$.  The reason for working with both $a/T$ and $a/s^{1/3}$ is that we wish to compare the drag force in the anisotropic plasma to that in the isotropic plasma, and this can be done at least in two different ways: the two plasmas can be taken to have the same temperatures but different entropy densities, or the same entropy densities but different temperatures. 

The drag force $F(v,\vp,a/T)$ in units of the isotropic drag force in a plasma at the same temperature is shown in Fig.~\ref{3Dtemp}. The drag force $F(v,\vp,a/s^{1/3})$ in units of the isotropic drag force in a plasma at the same entropy density is shown in Fig.~\ref{3Dentro}.  With a few exceptions, the results are qualitatively similar. In both cases we see that the anisotropic drag is larger than the isotropic drag except in a region near the $x$-axis. This region is more clearly shown in Fig.~\ref{inter}: the curves in that figure are  the intersections between the two surfaces shown in each of the corresponding 3D plot in Figs.~\ref{3Dtemp} or Figs.~\ref{3Dentro}.  Considering that the value of $a/T$ varies by a factor of 62 between  the top and the bottom curves in Fig.~\ref{inter}, we see that the region in question depends relatively mildly on the magnitude of the anisotropy.

For motion along the longitudinal $z$-direction, the anisotropic drag is greater than the isotropic drag for any value of $v$. For any direction of motion $\vp\neq \pi/2$, the ratio $F_\mt{aniso}/F_\mt{iso}$ diverges as $1/\sqrt{1-v^2}$ in the ultra-relativistic limit $v\to 1$ irrespectively of whether the comparison is made at the same temperature or at the same entropy density, as we prove analytically in Appendix \ref{ultra}.\footnote{We recall that we first send the quark mass to infinity and then $v\to 1$. See the penultimate paragraph of Sec.~\ref{intro}.} 
In other words, for motion not perfectly aligned with the transverse $x$-direction, the anisotropic drag becomes arbitrarily larger than the isotropic one as the ultra-relativistic limit is approached. This is most clearly illustrated in Fig.~\ref{dragvelo}, which shows constant-$\vp$ slices of the (c) and (d) plots in Figs.~\ref{3Dtemp} and Figs.~\ref{3Dentro}. We will come back to this result in Sec.~\ref{discussion}. 

\begin{figure}
\begin{center}
\begin{tabular}{cc}
\setlength{\unitlength}{1cm}
\includegraphics[width=6cm,height=4cm]{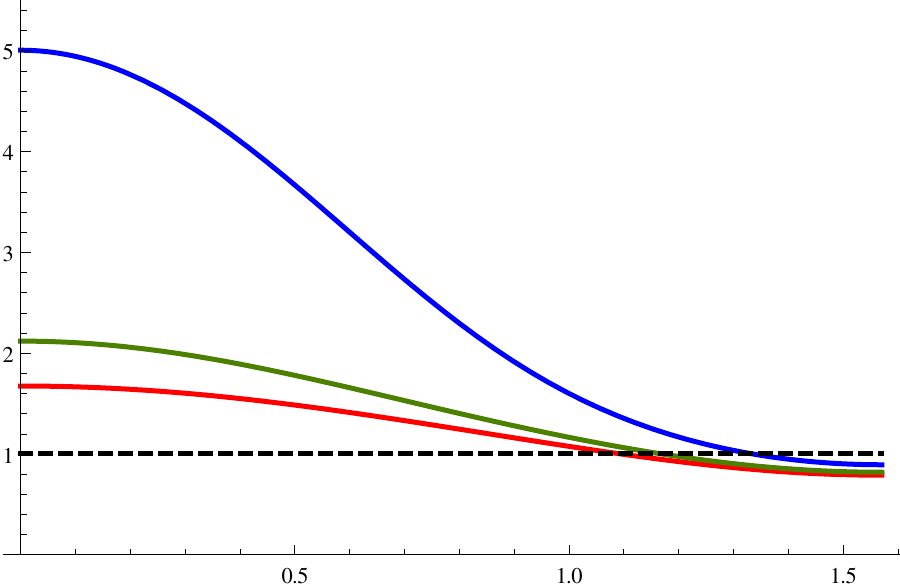} 
\qquad & \qquad\qquad
\includegraphics[width=6cm,height=4cm]
{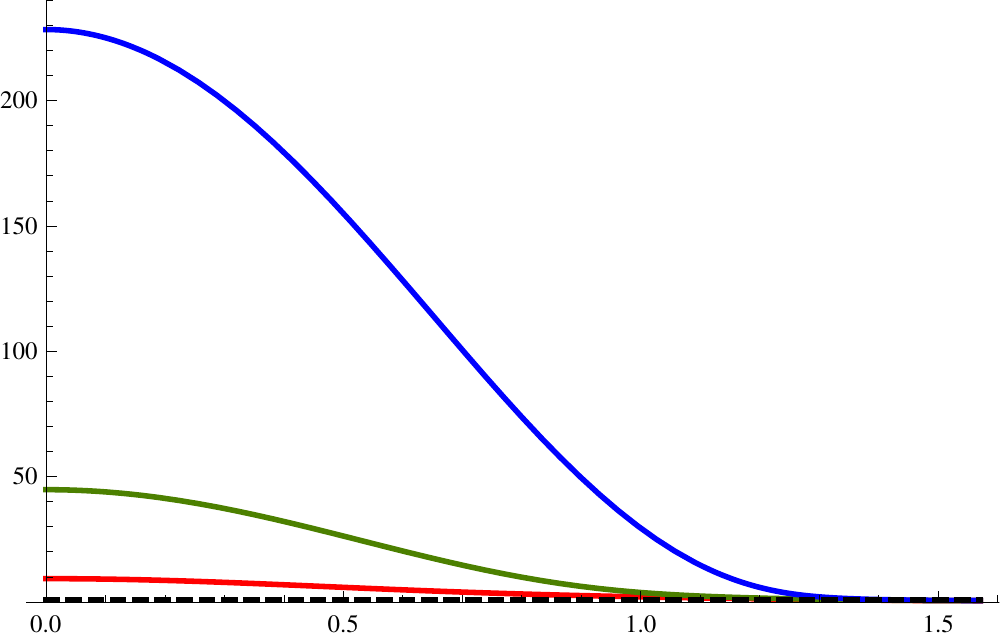}
\put(-420,45){\rotatebox{90}{$F/F_\mt{iso}(T)$}} 
\put(-190,45){\rotatebox{90}{$F/F_\mt{iso}(T)$}} 
 \\
$\vp$ & \qquad\qquad $\vp$ \\
& \\
\includegraphics[width=6cm,height=4cm]{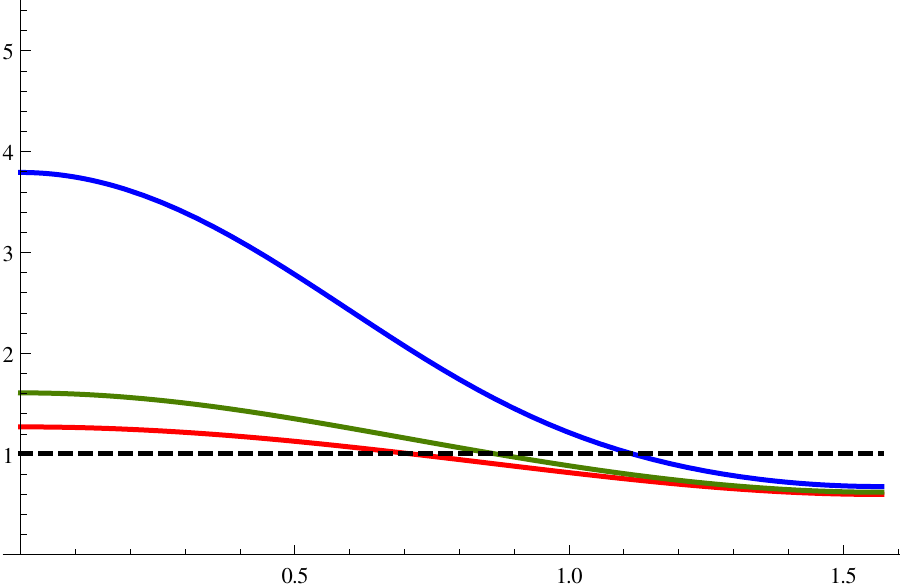} 
\qquad & \qquad\qquad
\includegraphics[width=6cm,height=4cm]
{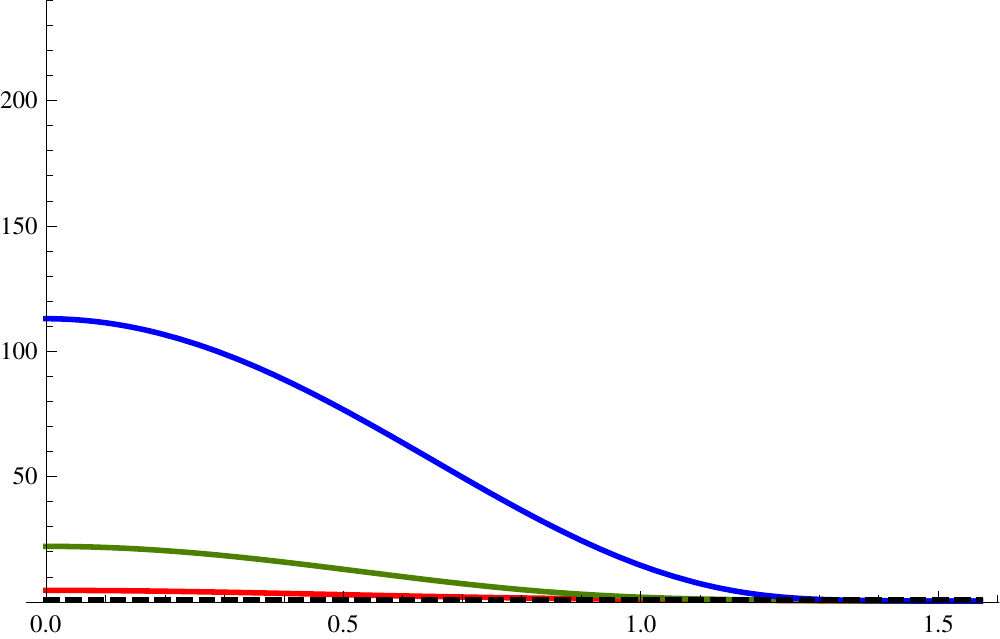}
\put(-420,43){\rotatebox{90}{$F/F_\mt{iso}(s)$}} 
\put(-190,43){\rotatebox{90}{$F/F_\mt{iso}(s)$}} 
 \\
$\vp$ & \qquad\qquad $\vp$ 
\end{tabular}
\end{center}
\caption{Drag force as a function of the direction of motion $\vp$, measured with respect to the longitudinal direction $z$, for a quark moving  through an anisotropic plasma with $a/T=12.2$, or equivalently $a \nc^{2/3}/s^{1/3}=6.24$,  (left column) and $a/T=86$, or equivalently $a \nc^{2/3}/s^{1/3}=35.5$,  (right column) at three different velocities (curves from top to bottom) $v= 0.9, 0.7, 0.5$. $F$ is plotted in the appropriate units to facilitate comparison with the isotropic result for a plasma at the same temperature (top row) or at the same entropy density (bottom row). The isotropic result is given in eqs.~\eqn{fisoT} and \eqn{fisos}.
}
\label{dragangle}
\end{figure}
The ratio $F_\mt{aniso}/F_\mt{iso}$ is always finite for motion along the transverse $x$-direction. (Incidentally,  this implies that the limits $\vp \to \pi/2$ and $v\to 1$ do not commute.) In this case we must distinguish between the comparisons at  equal  temperature or at equal entropy density. In the first case, our numerical results indicate that the anisotropic drag is smaller than the isotropic one for $0\leq v  < v_c$ and larger than the isotropic one for $v_c < v \leq 1$, and we have confirmed this analytically in the  limits of small and large anisotropies (see the Appendices). The velocity $v_c$ at which the transition takes place is $v_c \simeq 0.9$ for small anisotropies and it approaches 1 as the anisotropy increases.

In the second case our numerical results indicate that the anisotropic drag is smaller than the isotropic one for all $v\in [0,1]$ provided $a/s^{1/3}$ is small enough. In the opposite limit, $a/s^{1/3} \gg 1$, the anisotropic drag stays smaller than the isotropic one for small velocities and becomes larger above some critical velocity. We have confirmed this  analytically in the Appendices.  For a fixed $v$, the angle with respect to the $z$-direction beyond which the anisotropic drag may become smaller than the isotropic drag is shown in the constant-$v$ slices of Fig.~\ref{dragangle}.

\begin{figure}
\begin{center}
\begin{tabular}{cc}
\setlength{\unitlength}{1cm}
\includegraphics[width=6cm,height=4cm]{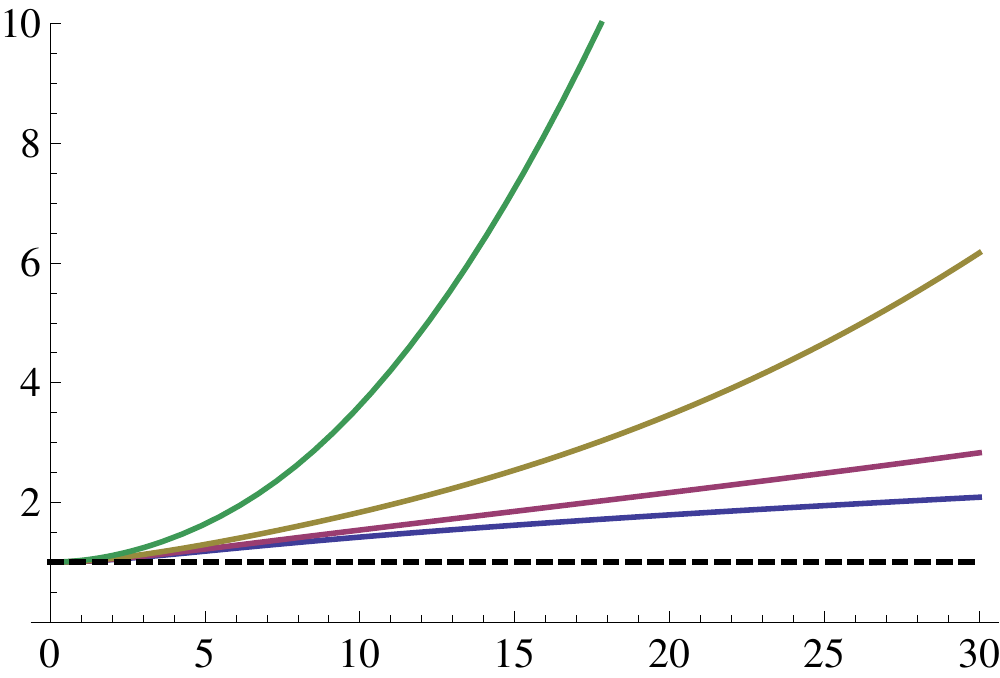} \qquad & \qquad\qquad
\includegraphics[width=6cm,height=4cm]{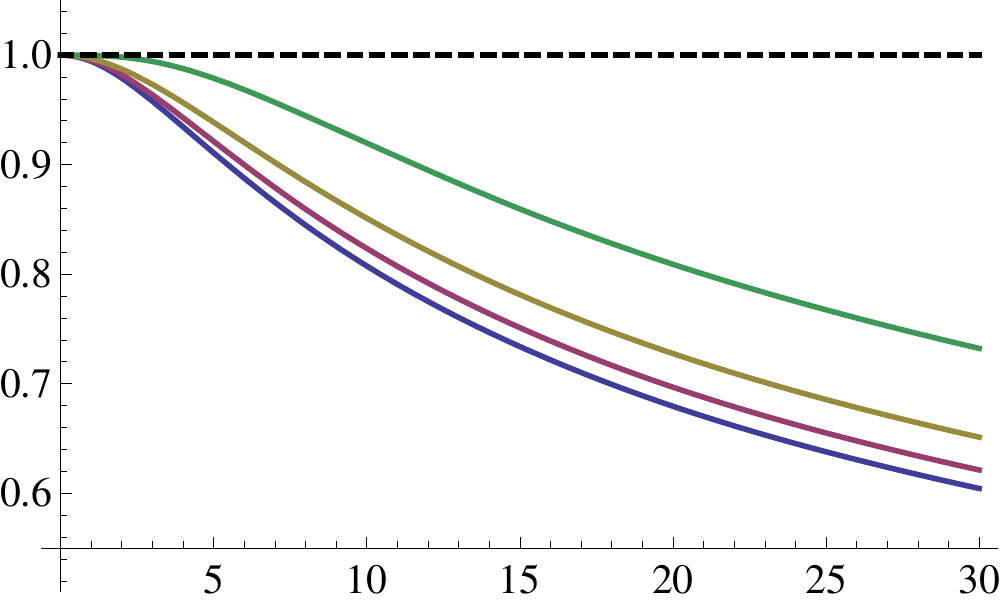}
\put(-420,45){\rotatebox{90}{$F/F_\mt{iso}(T)$}} 
\put(-190,45){\rotatebox{90}{$F/F_\mt{iso}(T)$}} 
 \\
$a/T$ & \qquad\qquad $a/T$ \\
& \\
\includegraphics[width=6cm,height=4cm]{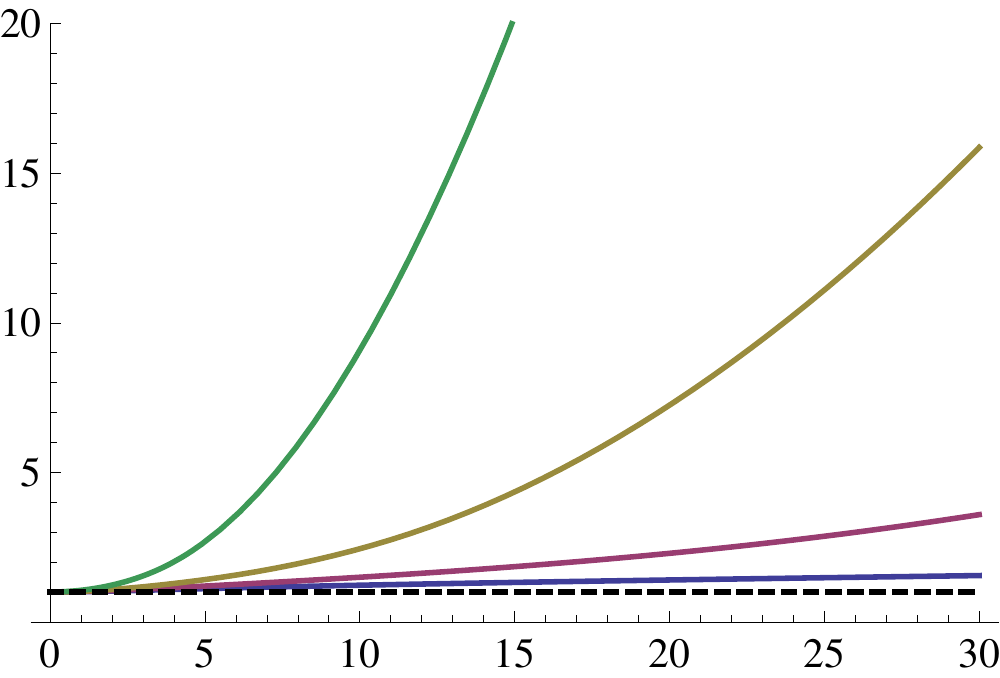} \qquad & \qquad\qquad
\includegraphics[width=6cm,height=4cm]{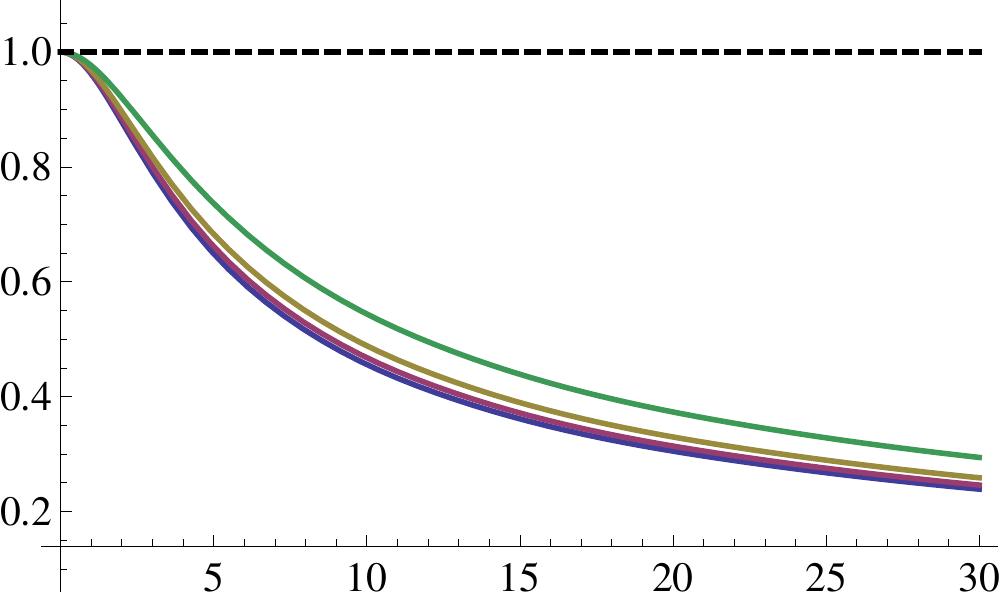}
\put(-420,43){\rotatebox{90}{$F/F_\mt{iso}(s)$}} 
\put(-190,43){\rotatebox{90}{$F/F_\mt{iso}(s)$}} 
 \\
$a\nc^{2/3}/s^{1/3}$ & \qquad\qquad $a\nc^{2/3}/s^{1/3}$ 
\end{tabular}
\end{center}
\caption{Drag force as a function of the anisotropy for a quark moving along the longitudinal $z$-direction, i.e.~at $\vp=0$ (left column) or along the transverse $x$-direction, i.e.~at $\vp=\pi/2$ (right column), at four different velocities (curves from top to bottom) $v=0.9, 0.7, 0.5, 0.25$. $F$ and $a$ are plotted in the appropriate units to facilitate comparison with the isotropic result for a plasma at the same temperature (top row) or at the same entropy density (bottom row). The isotropic result is given in eqs.~\eqn{fisoT} and \eqn{fisos}.
}
\label{dragani1}
\end{figure}

\begin{figure}
\begin{center}
\begin{tabular}{cc}
\setlength{\unitlength}{1cm}
\includegraphics[width=6cm,height=4cm]{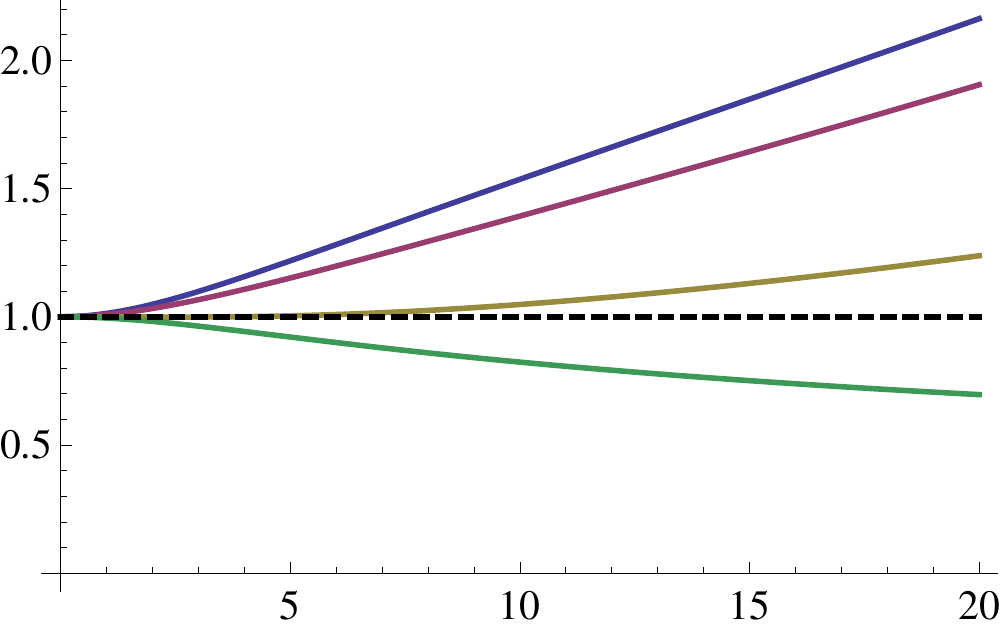} \qquad & \qquad\qquad
\includegraphics[width=6cm,height=4cm]{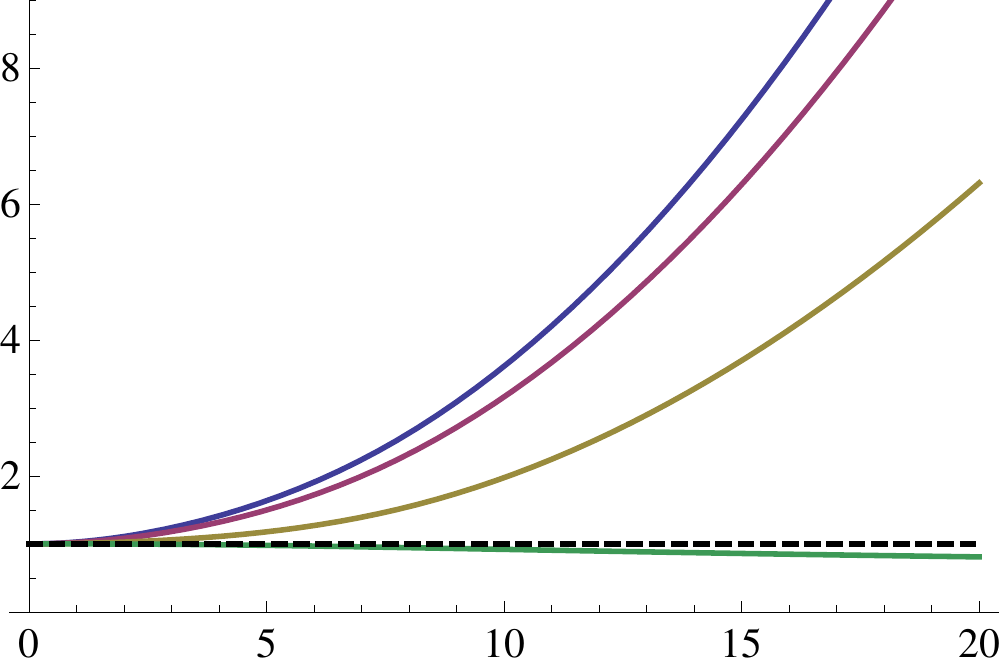}
\put(-420,45){\rotatebox{90}{$F/F_\mt{iso}(T)$}} 
\put(-190,45){\rotatebox{90}{$F/F_\mt{iso}(T)$}} 
 \\
$a/T$ & \qquad\qquad $a/T$ \\
& \\
\includegraphics[width=6cm,height=4cm]{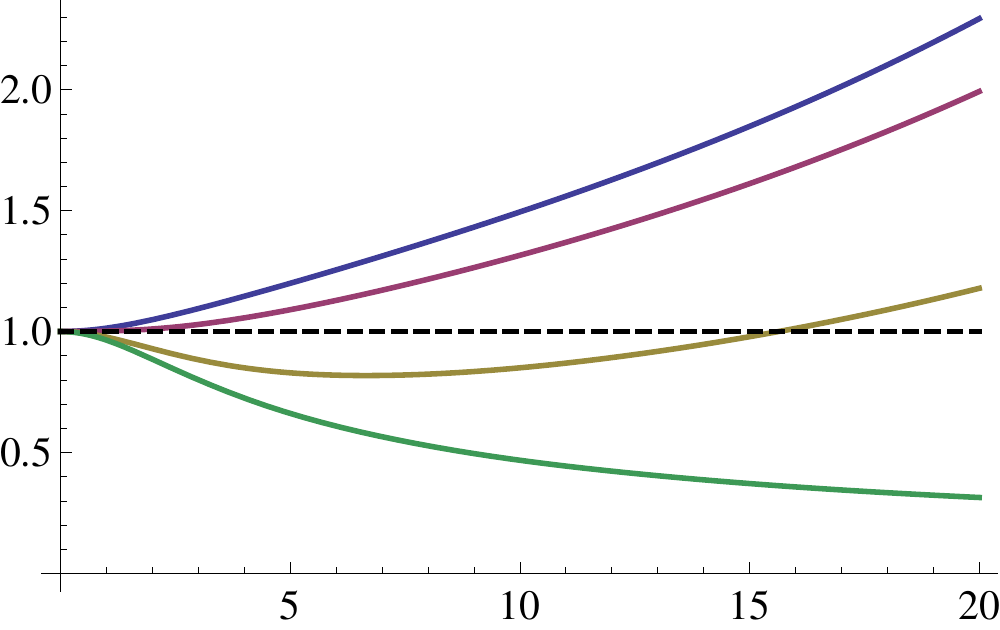} \qquad & \qquad\qquad
\includegraphics[width=6cm,height=4cm]{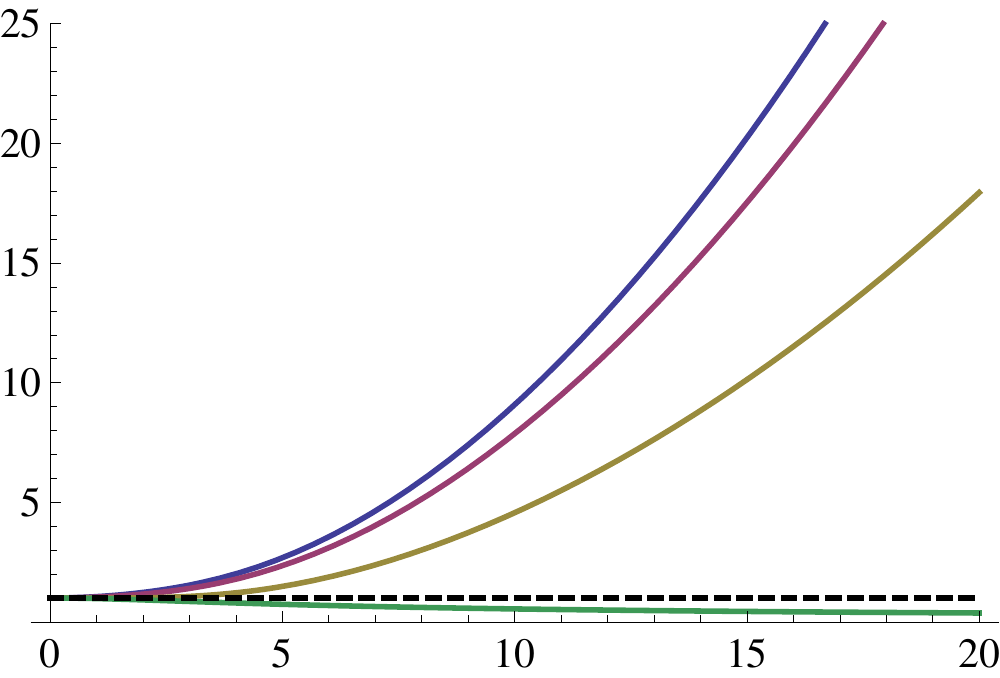}
\put(-420,43){\rotatebox{90}{$F/F_\mt{iso}(s)$}} 
\put(-190,43){\rotatebox{90}{$F/F_\mt{iso}(s)$}} 
 \\
$a\nc^{2/3}/s^{1/3}$ & \qquad\qquad $a\nc^{2/3}/s^{1/3}$ 
\end{tabular}
\end{center}
\caption{Drag force as a function of the anisotropy for a quark moving at $v=0.5$ (left column) or at $v=0.9$ (right column) along four different angles (curves from top to bottom) $\vp=0, \pi/6, \pi/3, \pi/2$ with respect to the longitudinal direction $z$. $F$ and $a$ are plotted in the appropriate units to facilitate comparison with the isotropic result for a plasma at the same temperature (top row) or at the same entropy density (bottom row). The isotropic result is given in eqs.~\eqn{fisoT} and \eqn{fisos}.
}
\label{dragani2}
\end{figure}

The dependence of the drag force on the anisotropy for fixed velocity is most clearly seen in Figs.~\ref{dragani1} and Figs.~\ref{dragani2}, where the ratio $F/F_\mt{iso}$ is plotted for several values of $v$ and $\vp$.

\begin{figure}[htb]
\begin{center}
\setlength{\unitlength}{1cm}
\includegraphics[width=6cm,height=4cm]{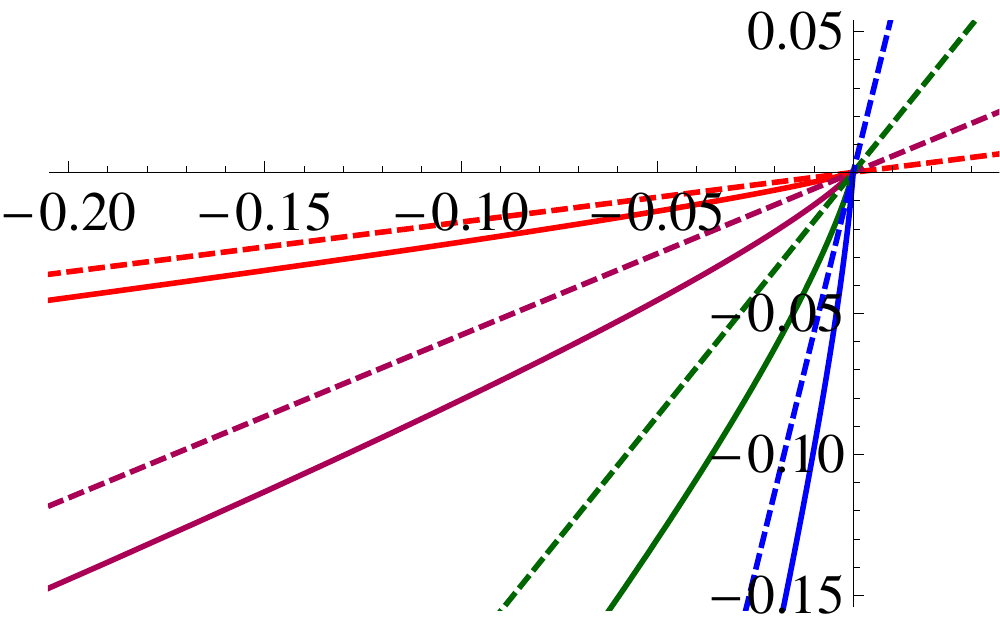}\qquad\qquad
\includegraphics[width=6cm,height=4cm]{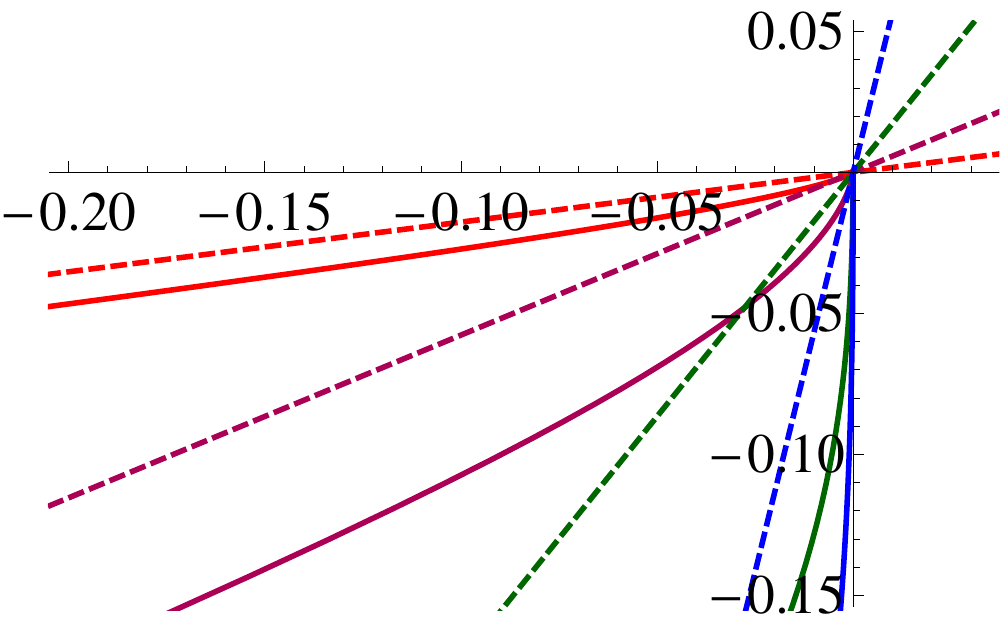}\qquad
 \begin{picture}(0,0)
   \put(-0.5,2.8){\small{$x$}}
   \put(-1.85,4.2){\small{$z$}}
    \put(-8.1,2.8){\small{$x$}}
   \put(-9.4,4.2){\small{$z$}}
 \end{picture}
\caption{Projection of a $t=0$ snapshot of the string profile \eqn{pro} (continuous curves) onto the $xz$-plane for a quark moving with velocity $v=0.7$ in four different directions (indicated by the dashed straight lines) that lie at angles (clockwise) $\vp=\pi/18, \pi/6, \pi/3, 8\pi/18$ with respect to the $z$-direction. The quark moves through  a plasma with anisotropy $a=12.2\, T$ (left) and $a=86\, T$ (right). The origin $(x,z)=(0,0)$ corresponds to the string endpoint, which lies at the boundary $u=0$. The coordinate $u$ increases along the curves away from this point.  The string curves in the $xz$-plane and (unless $\vp=0$ or $\pi/2$) only aligns itself with the velocity in the far infrared, i.e.~at large $u$.
}
\label{profile}
\end{center}
\end{figure}

In order to illustrate the geometric properties of the string solution, in Fig.~\ref{profile}  we have plotted the projection of the string profile onto the gauge theory directions. As anticipated, we see that the string curves in the $xz$-plane and (unless $\vp=0$ or $\pi/2$) only aligns itself with the velocity in the far infrared, i.e.~at large $u$. The  misalignment between the velocity $\vec v$, the drag force $\vec F$, and the tangent to the string profile at the string's endpoint $\vec \tau$ are shown in Fig.~\ref{tangents}. We see that, generally speaking, the misalignment becomes larger for larger anisotropies. This is more clearly quantified in Figs.~\ref{angleprofile}  and \ref{angledrag}, where the angles with respect to the $z$-direction of the tangent vector to the string and of the force are shown as a function of the angle of the direction of motion. From  Fig.~\ref{angleprofile} we see that the tangent vector to the string systematically `lags behind' the direction of motion as the latter varies from being aligned with the $z$-direction to being aligned with the $x$-direction. Only in these two limits does the string profile align itself entirely with the velocity. Moreover, the larger the anisotropy the more the string `wants' to stay aligned with the $z$-direction, changing direction quickly only as $\vp$ approaches $\pi/2$. From Fig.~\ref{angledrag} we see that the behaviour of the force is similar, except that for sufficiently large anisotropies its direction does not vary monotonically with the direction of the velocity. 

In order to gain an intuitive understanding of these geometric facts it is useful to think of the string in our anisotropic background \eqn{sol2} as a fishing string immersed in a river. Since the string provides a semiclassical description of the quark and its gluon cloud in the dual plasma, each of the statements below can be easily translated into gauge theory language. In the river analogy, the direction of the river's current  provides the anisotropic direction, and the fact that the anisotropy function $\ch (u)$ in \eqn{sol2} depends on the radial coordinate can be modeled by imagining that the magnitude of the  current depends on the depth. Under these circumstances it is clear that the string will curve as it descends deeper and deeper, since pieces of the string at different depths experience different degrees of anisotropy. It is also clear that each bit of the string deposits momentum into the river in a different direction that depends on the bit's  local orientation. The direction of the total (rate of) momentum deposition is  a combination of all of these contributions, and this combination equals the external force. It is thus clear that the external force will  not point in the same direction as the  vector tangent to the string at a generic point, in particular at its endpoint. Finally, the fact that the string eventually aligns with the velocity deep in the infrared can be understood as a consequence of the fact that the string `piles up' on top of the horizon of \eqn{sol2}. In the river's analogy, this could perhaps be modeled by imagining that the current vanishes at the bottom of the river, and that the string piles up there. To understand this point, note that a constant-$u$ slice of the metric \eqn{sol2} is locally isotropic, since the factor $\ch(u)$ can be locally absorbed through a rescaling of the $z$-coordinate. For generic $u$ this is irrelevant since the local isotropy is only experienced by an infinitesimal  bit of string. However, an infinite length of string lies between $\uh$ and $\uh + \epsilon$ for any $\epsilon >0$. Since this infinite piece of string experiences an effectively isotropic metric, it is not surprising that it aligns with the velocity of the quark, as it happens in the completely isotropic case  \cite{drag1,drag2}. 

We stress that the heuristic analogy above is only meant to provide a somewhat intuitive understanding of the geometric features described by Figs.~\ref{profile}-\ref{angledrag}, which arise rigorously  from the  minimization of the string action in our  anisotropic background \eqn{sol2}. In particular, we emphasize that, although it may seem counterintuitive at first sight,  there is no reason to expect the tangent vector to the string, the velocity and the force to be mutually aligned in the presence of an anisotropic medium. 

\begin{figure}[htb]
\begin{center}
\setlength{\unitlength}{1cm}
\includegraphics[width=6cm,height=4cm]
{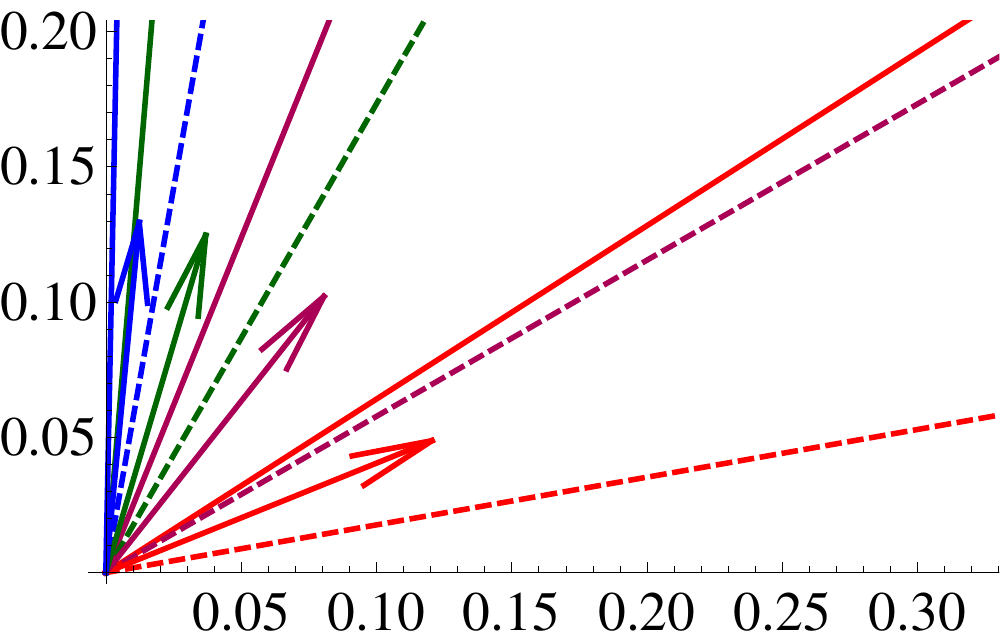}\qquad\qquad
\includegraphics[width=6cm,height=4cm]
{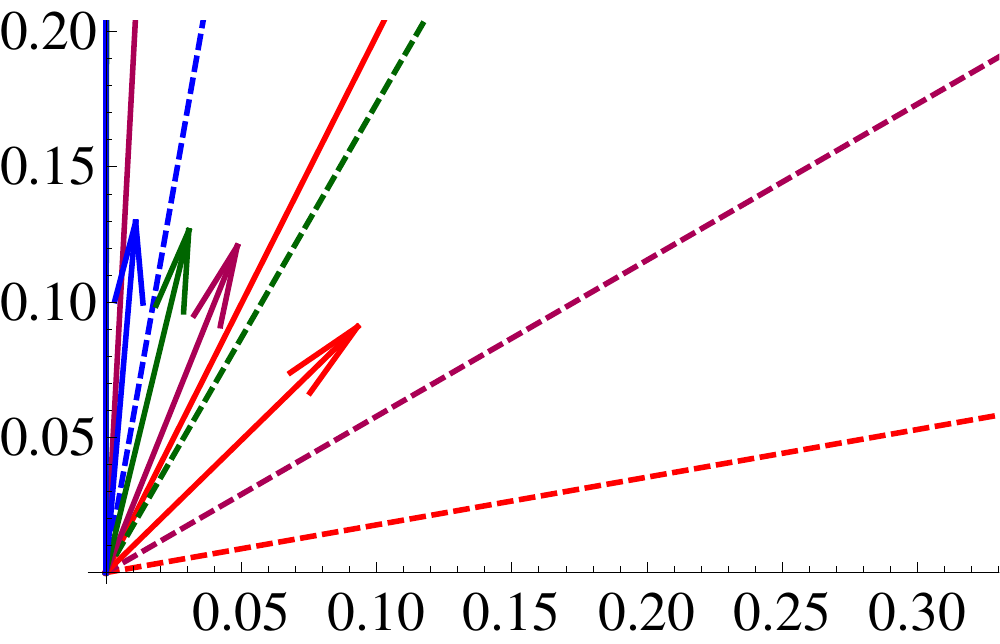}\qquad
 \begin{picture}(0,0)
   \put(-0.55,0.2){\small{$x$}}
   \put(-6.4,4.2){\small{$z$}}
    \put(-8.1,0.2){\small{$x$}}
   \put(-13.9,4.2){\small{$z$}}
 \end{picture}
\caption{Generic misalignment between the direction of the quark velocity (dashed straight lines), the direction of the force (arrows) and the direction tangent to the string profile at its endpoint (continuous straight lines). The quark velocity is $v=0.7$, its direction lies at  angles (clockwise) $\vp=\pi/18, \pi/6, \pi/3, 8\pi/18$ with respect to the $z$-direction, and the anisotropy is $a=12.2\, T$ (left) and $a=86\, T$ (right).
}
\label{tangents}
\end{center}
\end{figure}
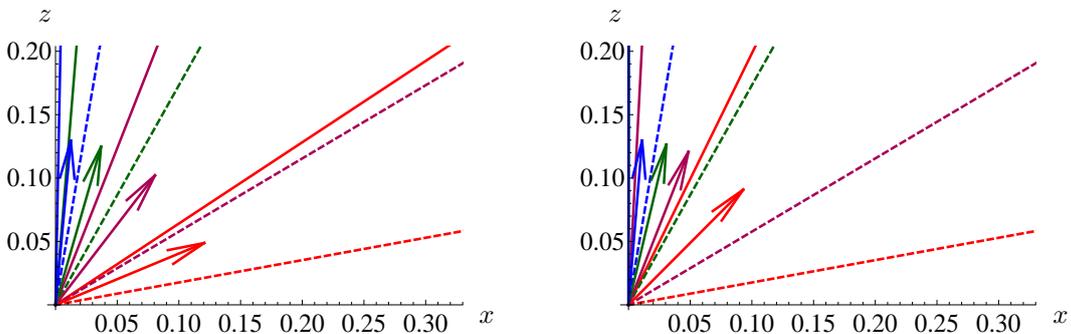
\begin{figure}[htb]
\begin{center}
\setlength{\unitlength}{1cm}
\includegraphics[width=6cm,height=4cm]{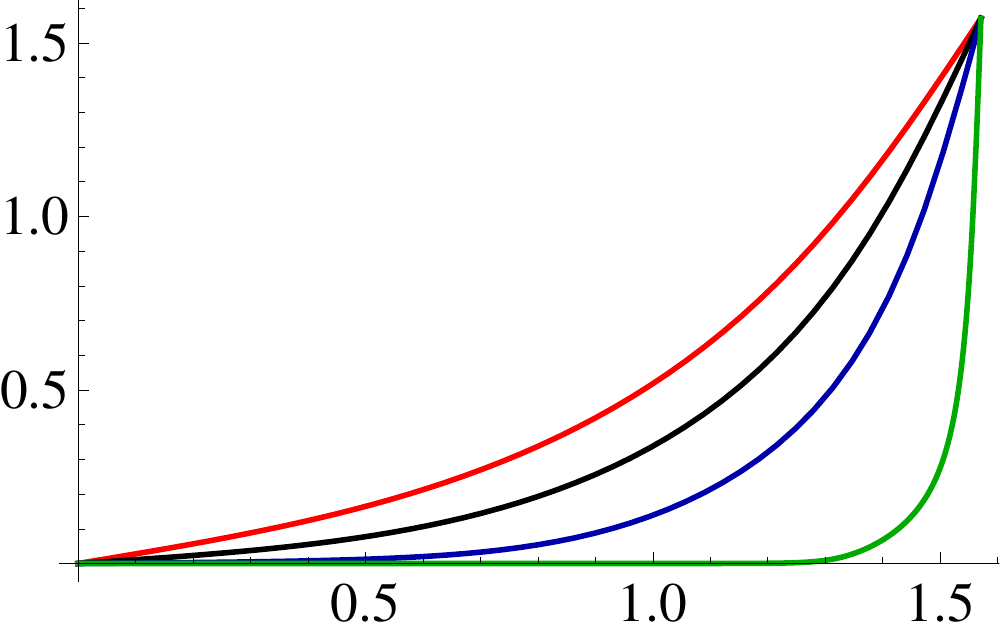}\qquad\qquad
\includegraphics[width=6cm,height=4cm]{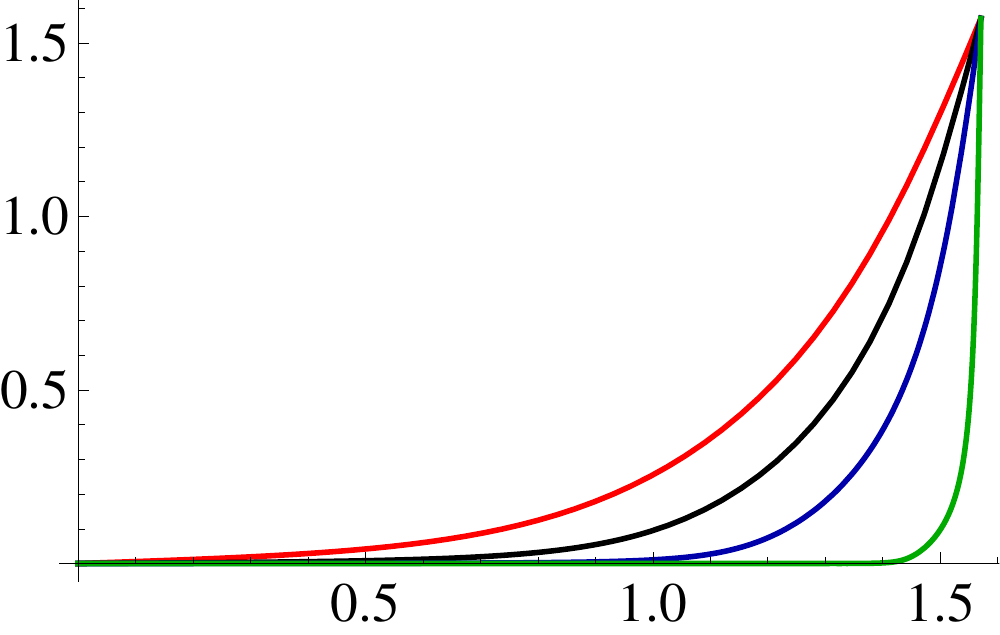}\qquad
 \begin{picture}(0,0)
   \put(-0.6,0.2){{$\vp$}}
   \put(-6.7 ,4.25){{$\vp_\tau$}}
    \put(-8.15,0.2){{$\vp$}}
   \put(-14.2,4.25){{$\vp_\tau$}}
 \end{picture}
\caption{Tendency of the string to align itself with the longitudinal direction $z$ for four different anisotropies (from top to bottom) $a/T = 12.2, 20.3,
42.6, 744$. The angle $\vp_\tau$ is the angle between the $z$-axis and the tangent vector to the string at its endpoint, defined as in eqn.~\eqn{phitau}. The angle $\vp$ is the angle between the $z$-axis and the velocity. The magnitude of the velocity is $v=0.7$ (left) and $v=0.9$ (right).
}
\label{angleprofile}
\end{center}
\end{figure}
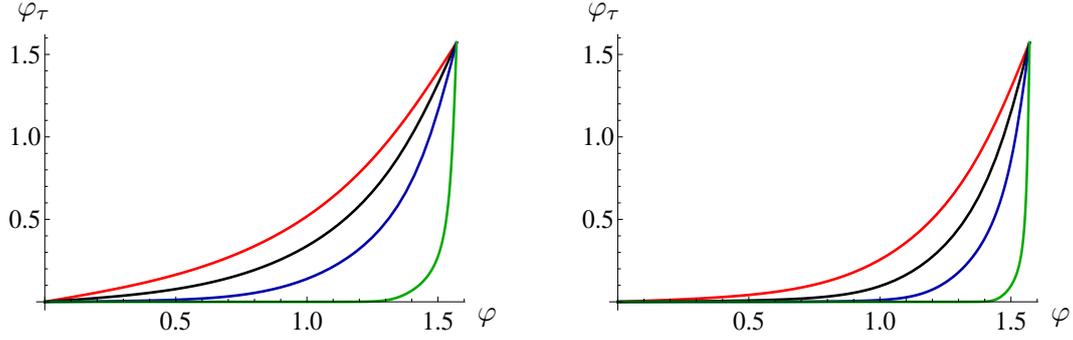
\begin{figure}[htb]
\begin{center}
\setlength{\unitlength}{1cm}
\includegraphics[width=6cm,height=4cm]{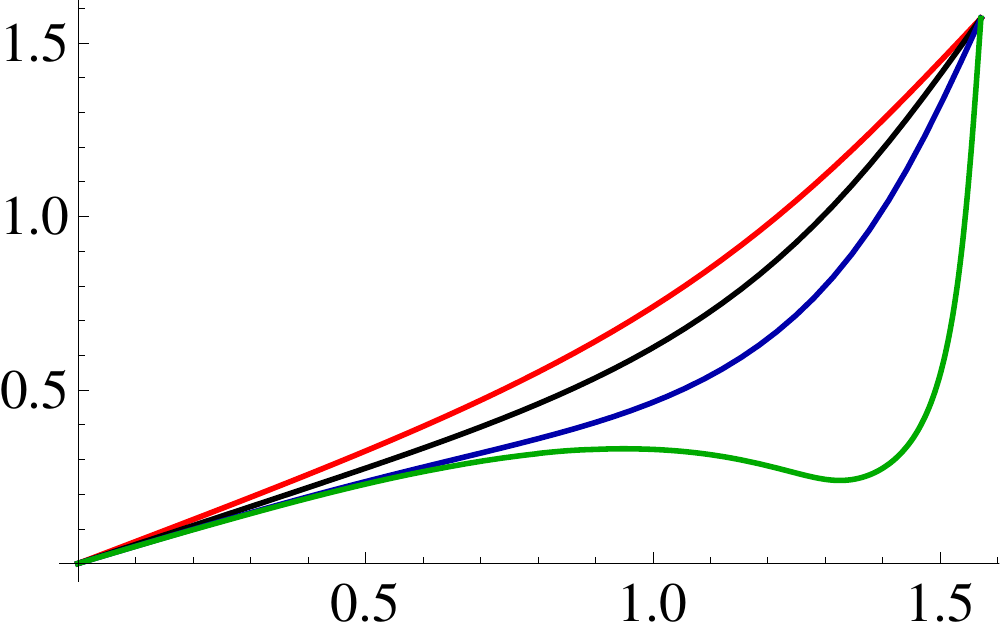}\qquad\qquad
\includegraphics[width=6cm,height=4cm]{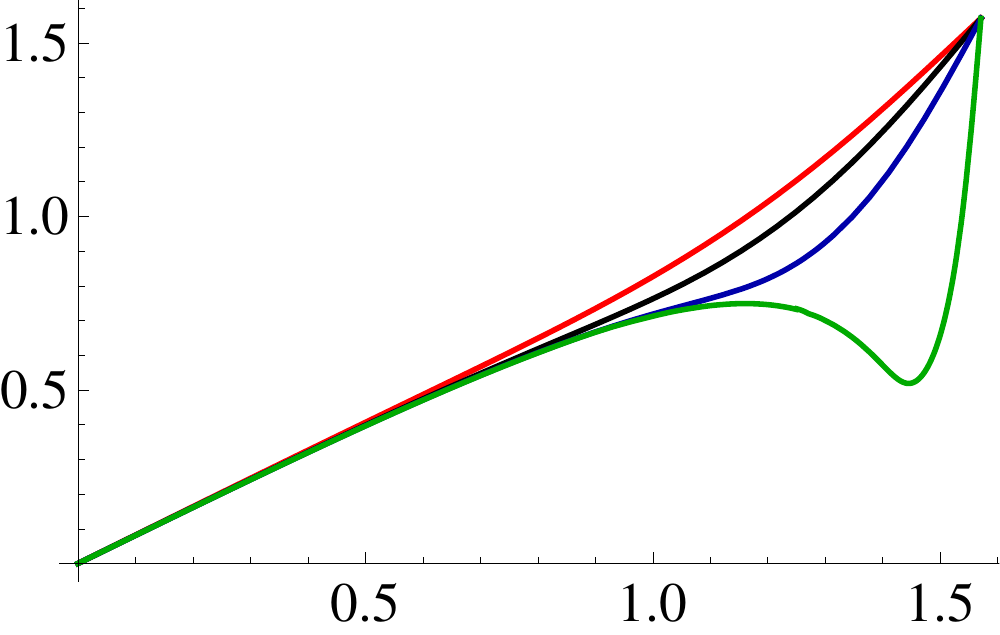}\qquad
 \begin{picture}(0,0)
   \put(-0.6,0.2){{$\vp$}}
  \put(-6.7 ,4.25){{$\vp_\mt{F}$}}
    \put(-8.15,0.2){{$\vp$}}
   \put(-14.2,4.25){{$\vp_\mt{F}$}}
 \end{picture}
\caption{Correlation between the direction of the force, $\vp_\mt{F}$, and the direction of the velocity, $\vp$ (with both angles measured with respect to the longitudinal direction $z$), for four different anisotropies (from top to bottom) $a/T = 12.2, 20.3, 42.6, 744$. 
The magnitude of the velocity is $v=0.7$ (left) and $v=0.9$ (right). 
}
\label{angledrag}
\end{center}
\end{figure}

\section{Discussion}  
\label{discussion}
We have analyzed the drag force exerted on an infinitely massive quark moving through an anisotropic ${\cal N}=4$ super Yang-Mills plasma described by the metric \eqn{sol2}. In this case the anisotropy is induced by a position-dependent theta term in the gauge theory, or equivalently by a position-dependent  axion on the gravity side. One may therefore wonder how sensitive the conclusions may be to the specific source of the anisotropy. In this respect it is useful to note that the gravity calculation involves only the coupling of the string to the background metric. This means that any anisotropy that gives rise to a qualitatively similar metric (and no Neveu-Schwarz $B$-field) will yield qualitatively similar results for the drag force irrespectively of the form of the rest of the supergravity fields. 

An example of a rather robust conclusion is the ultra-relativistic behaviour of the drag force.\footnote{We recall that we first send the quark mass to infinity and then $v\to 1$. See the penultimate paragraph of Sec.~\ref{intro}.}  We have seen that  the anisotropic solution \eqn{sol2} yields a drag force that  becomes arbitrarily larger than the isotropic one for all ultra-relativistic quarks except for those whose velocity is perfectly aligned with the transverse $xy$-plane. This follows from the fact that the near-boundary fall-off of the metric \eqn{sol2} takes the schematic form
\be
g_{\mu\nu} = \frac{L^2}{u^2} \left( \eta_{\mu\nu} + 
u^2 g^{(2)}_{\mu\nu} + u^4 g^{(4)}_{\mu\nu} + \cdots
\right) \,.
\label{near}
\ee
As $v$ grows closer and closer to 1 the string worldsheet develops a horizon closer and closer to the AdS boundary at $u=0$. As a consequence the physics in this limit is solely controlled by the near-boundary behaviour of the metric. Eq.~\eqn{hc} can then be solved using the asymptotic form \eqn{near} of the metric functions. Generically the solution to leading order is determined by the ${\cal O}(u^2)$-terms and yields $u_c^2 \propto 1-v^2$. Substituting in \eqn{dragv} one gets $F\propto 1/(1-v^2)$, or equivalently $F = \mu p$ with a momentum-dependent drag coefficient $\mu \propto p$.
For example, we show in  Appendix \ref{ultra} that the metric \eqn{sol2} yields a drag coefficient 
\be
\mu(p) \simeq \frac{\sqrt{\lambda} \, a^2 \cos^2 \vp}{8\pi M^2} \, p
\ee
at large $p$, where $M$ is the quark mass. In contrast, in the isotropic case of \cite{drag1,drag2} the ${\cal O}(u^2)$-terms in the metric are absent. This means that the solution of eq.~\eqn{hc} is $u_c^4 \propto 1-v^2$ and hence that in this case the drag force in the ultra-relativistic limit has a softer divergence $F_\mt{iso} \propto 1/\sqrt{1-v^2}$. Rewriting this in terms of the momentum gives $F =\mu p$ with $\mu$ a momentum-independent constant in this case. For certain choices of the parameters (for example for $\vp=\pi/2$ in our case) the ${\cal O}(u^2)$-terms in eq.~\eqn{hc} may vanish, in which case $F \propto 1/\sqrt{1-v^2}$. 

The above discussion makes it clear that the linear behaviour of the drag coefficient in the ultra-relavistic limit, 
$\mu \propto p$, depends  solely on two features of the solution: The presence of the $g^{(2)}_{\mu\nu}$ term in the near-boundary expansion of the metric, and the fact that the metric \eqn{near} be non-boost-invariant at order $u^2$ (i.e.~that $g^{(2)}_{\mu\nu}$ not be proportional to $\eta_{\mu\nu}$). The latter condition is necessary because otherwise there would be no solution for $u_c$ at order $u^2$. Note that adding temperature to an otherwise boost-invariant metric  will only affect $g^{(4)}_{\mu\nu}$, and thus this is not enough to make $g^{(2)}_{\mu\nu}$ non-boost-invariant. This conclusion is consistent with the fact that $g^{(2)}_{\mu\nu}$ is only a function of the external sources which the theory is coupled to.

Interesting backgrounds with non-zero $g^{(2)}_{\mu\nu}$ include bona fide string theory constructions, i.e.~smooth supergravity solutions with a well known gauge theory dual, as well as  `ad hoc' backgrounds, i.e.~backgrounds that do not solve supergravity equations but are phenomenologically motivated. An example in the first category is the supergravity flow \cite{star0,star1,star2} dual to the ${\cal N}=2^*$  deformation of the ${\cal N}=4$ super Yang-Mills theory by fermion (and scalar) masses.  An example in the second category is the linear-dilaton background of Refs.~\cite{KKSS,KKSS2,nonconf1}. Both sets of examples have in common that, at zero temperature, conformal invariance is broken but the full Lorentz symmetry of the boundary theory is preserved. The breaking of conformality results in a momentum-dependent drag coefficient $\mu(p)$, as shown  in \cite{nonconf4} for the ${\cal N}=2^*$ theory, and in \cite{nonconf1,nonconf3} for the linear-dilaton background. However, in both sets of examples $g^{(2)}_{\mu\nu}$ is boost-invariant, since this term (unlike $g^{(4)}_{\mu\nu}$) is unaffected by the further breaking of conformal symmetry that occurs at non-zero temperature. As a consequence, in the ultra-relativistic limit
the drag-coefficient becomes momentum-independent and approaches a constant. Thus, as measured by this particular observable, one may regard the breaking of conformality in the anisotropic plasma of \cite{prl,jhep} as more severe than in the backgrounds above.

\begin{acknowledgments}
It is a pleasure to thank Roberto Emparan, Tomeu Fiol, Alberto G\"uijosa, and specially Jorge Casalderrey-Solana and Ioannis Papadimitriou for discussions. MC is supported by a postdoctoral fellowship from Mexico's National Council of Science and Technology (CONACyT). 
We acknowledge financial support from 2009-SGR-168, MEC FPA2010-20807-C02-01, MEC FPA2010-20807-C02-02 and CPAN CSD2007-00042 Consolider-Ingenio 2010 (MC, DF and DM), and from DE-FG02-95ER40896 and CNPq (DT).
\end{acknowledgments}

\appendix

\section{Ultra-relativistic limit}
\label{ultra}
In the limit $v\to 1$ the value of $u_c$ that solves eqn.~\eqn{hc} approaches the boundary, i.e.~$u_c\to 0$.  Therefore in this limit $u_c$ can be determined from the near-boundary expansion of the metric functions, which takes the form: 
\bea
&& 
\cf = 1+\frac{11\, a^2}{24}u^2+\left({\cal F}_4
+\frac{7\, a^4}{12} \log u\right) u^4 +  {O}(u^6) \,, \nonumber \\
&& 
\cb = 1-\frac{11\, a^2}{24}u^2+\left({\cal B}_4
-\frac{7\, a^4}{12} \log u\right) u^4 +  {O}(u^6) \,, \nonumber \\
\cr && \ch =  1+\frac{a^2}{4}u^2
- \left(\frac{2\, {\cal B}_4}{7}-\frac{5\, a^4}{4032}-
\frac{a^4}{6} \log u \right)u^4 +  {O}(u^6) \,,
\label{expansionFunctions}
\eea
The coefficients $\cb_4, \cf_4$ depend on $a,T$ and are related to the energy and the pressures of the plasma eqs.~(35) in \cite{jhep}. They 
are not determined by the near-boundary analysis but must instead be read off from a full bulk solution. 

We must solve \eqn{hc} for $u_c$ to leading order in $1-v^2$. For $\cos \vp \neq 0$ we only need to consider the terms of ${\cal O}(u^2)$ in the metric functions, and the solution is
\be
\frac{1}{u_c^2} = \frac{a^2 \,v^2 \cos^2 \vp}{4(1-v^2)} \,. 
\qquad \qquad  [\cos \vp \neq 0]
\ee
Substituting into \eqn{dragv} we obtain the drag force
\be
\vec F = \frac{\sqrt{\lambda}}{8\pi} 
\, a^2 \cos^2 \vp \, \frac{v^3}{1-v^2} 
( \sin \vp, \cos \vp) \,.
\qquad \qquad [\cos \vp \neq 0] 
\ee
The divergence when $v\to 1$ contrasts with the softer behaviour \eqn{fisoT}-\eqn{fisos} of the isotropic case. We conclude that for $\cos \vp \neq 0$ the ratio $F_\mt{aniso}/F_\mt{iso}$ diverges in the limit $v\to 1$ as $1/\sqrt{1-v^2}$, in agreement with our numerical results displayed in Fig.~\ref{dragvelo}. Note that this is true even if the two plasmas have different temperatures and/or different entropy densities, since in the anisotropic case $F$ diverges as $1/(1-v^2)$ irrespectively of the temperature or the entropy density. 

The previous analysis shows that the limits $v\to 1$ and $\vp \to \pi/2$ do not commute. This is because if we first set $\cos \vp=0$ then the terms of order $u^2$ cancel out in eqn.~\eqn{hc} and we must go to order $u^4$. The solution in this case is 
\be
\frac{1}{u_c^2}= \frac{T^2}{\sqrt{1-v^2}}
\sqrt{\frac{121}{576}\frac{a^4}{T^4}- 
\frac{\cb_4+\cf_4}{T^4}}\,,
\qquad \qquad [\cos \vp = 0] 
\ee
which yields the drag force
\be
F_x = \frac{\sqrt{\lambda}\, T^2}{2\pi}  \frac{v}{\sqrt{1-v^2}}
\sqrt{\frac{121}{576}\frac{a^4}{T^4}- 
\frac{\cb_4+\cf_4}{T^4}}\,.
\qquad \qquad [\cos \vp = 0] 
\label{fx} 
\ee
This result is valid for any value of $a/T$, large or small, and it implies that the ratio $F_\mt{aniso}/F_\mt{iso}$ is finite in the limit $v\to 1$ and given by
\be
\frac{F_x}{F_\mt{iso}}=\frac1{\pi^2} \sqrt{\frac{121a^4}{576T^4}-
\left (\frac{\cf_4+\cb_4}{T^4} \right)} \,.
\qquad \qquad [\cos \vp = 0] 
\label{Fv1}
\ee 
This result is valid for any $a$, large or small (as long as the motion is exactly aligned with the $x$-direction). In order to proceed further we need  analytic expressions for the  coefficients $\cf_4, \cb_4$. These are known in the limiting cases of small and large $a/T$. In the first case they are given in eqn.~(175) of \cite{jhep}: 
\bea
{\cal F}_4 &=& -\pi^4 T^4 -\frac{9 \pi^2 T^2}{16}a^2
- \left[ \frac{101}{384}
-\frac{7}{12}\log \left(\frac{2\pi T}{a} \right)
-\frac{7}{12}\log \left(\frac{a}{\Lambda} \right)\right] a^4 +{O}(a^6)\,,\cr
{\cal B}_4&=& \frac{7\pi^2T^2}{16}a^2+\left[\frac{593}{1152}
-\frac{7}{12}\log \left(\frac{2\pi T}{a} \right)
-\frac{7}{12}\log \left(\frac{a}{\Lambda} \right)\right]a^4 +{O}(a^6)\,,
\label{F4B4_T}
 \eea
where $\Lambda$ is a reference scale related to the conformal  anomaly. Substituting into \eqn{Fv1} we find
\be
\frac{F_x}{F_\mt{iso}}= 1+ 
\frac{a^2}{16 \pi^2 T^2} + 
{\cal O}\left( \frac{a^4}{T^4} \right) \,.
\qquad \qquad [\cos \vp = 0, \mbox{small $a/T$}] 
\label{res}
\ee
Note that the dependence on the reference scale $\Lambda$ has cancelled out in this result, as expected from the discussion in the last paragraph of Sec.~\ref{grav}. The result \eqn{res} shows that the drag force on an ultra-relativistic quark moving  along the transverse directions in an anisotropic plasma with small $a/T$ is greater than the drag in an isotropic plasma at the same temperature, in agreement with our numerical results. In order to make this comparison at equal entropy densities we use the fact that the entropy density at small $a/T$ is given by (see eq.~(174) in \cite{jhep})
\bea
s=\frac{\pi^2  \nc^2   T^3}{2}+\frac{ \nc^2  T}{16}a^2
+{O}\left( \frac{a^4}{T} \right) \,.
\label{sss1}
\eea
Inverting this relation,
\be
T = \left( \frac{2}{\nc^2 \pi^2} \right)^{1/3} s^{1/3}
\left[ 1 -
\frac{1}{24} \left( \frac{\nc^2}{2 \pi} \right)^{2/3} 
\frac{a^2}{s^{2/3}} + 
{O}\left( \frac{a^4}{s^{4/3}} \right)\right] \,,
\ee
substituting in \eqn{fx} and taking the ratio with \eqn{fisos} we arrive at
\be
\frac{F_x}{F_\mt{iso}}= 1- \frac{1}{48} \left( \frac{\nc^2}{2 \pi} \right)^{2/3}  
\frac{a^2}{s^{2/3}} + 
{\cal O}\left( \frac{a^4}{s^{4/3}} \right) \,.
\qquad \qquad [\cos \vp = 0, \mbox{small $a^3/s$}] 
\label{res2}
\ee
We see that, in contrast to the case of equal temperatures, the drag in the anisotropic plasma is smaller if the comparison is made at equal entropy densities, again in agreement with our numerical results. 

In the limit of large $a/T$ the coefficients $\cf_4, \cb_4$ can be obtained by combining eqs.~(35), (89) and (90) of \cite{jhep}. The result is
\bea
\cf_4 &=& \frac{1}{132} \left[ 132 a^4 c_\mt{int} 
 +77 a^4 \log \left(\frac{a}{\Lambda} \right) 
-348 c_\mt{ent} \pi^2 a^{1/3} T^{11/3} + \cdots \right] , \\
\cb_4 &=& \frac{1}{6336} \left[ -6336 a^4 c_\mt{int} 
+ 1331 a^4 - 3696 a^4 \log \left(\frac{a}{\Lambda} \right) 
+4032 c_\mt{ent} \pi^2 a^{1/3} T^{11/3} + \cdots\right] ,
\,\,\,\,\,\,\,\,\,\,\,\,\,\,\,\,\,
\eea
where $c_\mt{int}$ in an integration constant and $c_\mt{ent}$ is the constant introduced in \eqn{larges}. Following the same procedure as in the small-$a$ case we find that the ratio at equal temperatures is 
\be
\frac{F_x}{F_\mt{iso}}= 
\frac{\sqrt{2 c_\mt{ent}}}{\pi} \, \frac{a^{1/3}}{T^{1/3}} + \cdots \,,
\qquad \qquad [\cos \vp = 0, \mbox{large $a/T$}] 
\label{res3}
\ee
where the dots stand for subleading terms in the large $a/T$ limit, and at equal entropy densities it is 
\be
\frac{F_x}{F_\mt{iso}}= 
\frac{1}{2^{1/6} \pi^{1/3} c_\mt{ent}^{3/16}}
\left( \frac{s}{\nc^2} \right)^{1/48} \frac{1}{a^{1/16}}
+ \cdots \,, 
\qquad \qquad [\cos \vp = 0, \mbox{large $a^3/s$}] 
\ee
We conclude that at large anisotropies the ultra-relativistic drag  in the anisotropic case is always greater than the isotropic drag.

\section{Small-anisotropy limit}
\label{analytical}
For small values of $a/T$ analytic expressions for the metric functions can be found \cite{prl,jhep} by perturbing around the isotropic case. The result is
\bea
\label{Fanalytic}
\cf (u)&=&1-\frac{u^4}{\uh^4}+a^2\mathcal{F}_2(u)
+\mathcal{O}(a^4) \,, \\
\label{Banalytic}
\cb (u)&=&1+a^2\mathcal{B}_2(u)+\mathcal{O}(a^4)\,, \\
\label{Hanalytic}
\log \ch(u)&=&
\frac{a^2 \uh^2}{4} \log\left[ 1 + \frac{u^2}{\uh^2} \right] +\mathcal{O}(a^4)\,,
\eea
where
\bea
\mathcal{F}_2(u)&=&\frac{1}{24\uh^2}\left[
8u^2(\uh^2-u^2)-10u^4\log 2+
3 \uh^4+7u^4\log \left( 1+\frac{u^2}{\uh^2} \right) \right] \,, \\
\mathcal{B}_2(u)&=&-\frac{\uh^2}{24} \left[
\frac{10u^2}{\uh^2+u^2}+\log\left( 1+\frac{u^2}{\uh^2}
\right) \right] \,.
\eea
Using these expressions in the general formulas of Sec.~\ref{dragsec} we obtain the correction to the isotropic result for the drag force at leading order in $a/T$. The result for the drag force along the longitudinal direction $z$ is 
\be
F_z= F_\mt{iso}(T)
\left[ 1 + \left( \frac{a^2}{T^2} \right)\,
\frac{1-v^2+\sqrt{1-v^2}+(1+v^2)\log \left( 1+\sqrt{1-v^2} \right)}
{24\pi^2(1-v^2)}
+\mathcal{O}\left( \frac{a^4}{T^4} \right) \right] \,,
\label{one}
\ee
whereas for the transverse direction $x$ it is
\be
F_x=
F_\mt{iso}(T)\left[ 1 + \left( \frac{a^2}{T^2} \right)\,
\frac{1-v^2+\sqrt{1-v^2}+(4 v^2-5)\log \left( 1+\sqrt{1-v^2} \right)}
{24\pi^2(1-v^2)}
 +\mathcal{O}\left( \frac{a^4}{T^4} \right) \right]\,,
\label{two}
\ee
The $\mathcal{O}(a^2/T^2)$ correction in \eqn{one} is positive for $v\in [0,1]$, whereas that in \eqn{two} is negative for $0\leq v<v_c$ and positive for $v<v_c\leq 1$, where $v_c \simeq 0.9$. This means that, for small enough an anisotropy,  the drag force along the longitudinal direction in the anisotropic plasma is always larger than the drag force in an isotropic plasma at the same temperature (but different entropy density). In the case of motion in the transverse direction the anisotropic drag is smaller than the isotropic drag for low $v$ and larger for high $v$. This is in agreement with the numerical results of Sec.~\ref{results}.

In order to compare with an isotropic plasma at the same entropy density (but different temperature) we use the relation found in \cite{jhep} for the entropy density of the anisotropic plasma:
\begin{equation}\label{sandT}
s=\frac{\pi^2N^2_cT^3}{2}
\left[ 1+ \frac{a^2}{8\pi^2 T^2} 
+\mathcal{O}\left( \frac{a^4}{T^4} \right)\right]\,.
\end{equation}
Inverting this relation and substituting in \eqn{one} and \eqn{two} we get:
\bea
F_z&=&F_\mt{iso}(s)
\left[  1 +   \frac{a^2}{24}\left(\frac{N^2_c}{2\pi s}\right)^{2/3} 
 \frac{\sqrt{1-v^2}-(1-v^2)+(1+v^2)\log ( 1+\sqrt{1-v^2} )}
{1-v^2}  + 
{ \mathcal{O} \left( \frac{a^4}{s^{4/3}} \right)} \right] \,, \nonumber \\
F_x&=&F_\mt{iso}(s)
\left[  1 +   
 \frac{a^2}{24}\left(\frac{N^2_c}{2\pi s}\right)^{2/3} 
 \frac{\sqrt{1-v^2}-(1-v^2)+(4 v^2-5)\log ( 1+\sqrt{1-v^2} )}
{1-v^2}  + 
{ \mathcal{O} \left( \frac{a^4}{s^{4/3}} \right)} \right] \,. \nonumber \\
\eea
In this case the leading correction is positive for all $v$ in $z$-direction and negative for all $v$ in the $x$-direction. Again, this is in agreement with the numerical results of Sec.~\ref{results}.

\end{document}